# Role of Ni substitution on structural, magnetic and electronic properties of epitaxial $CoCr_2O_4$ spinel thin films

P. Mohanty,[1,*] S. Chowdhury,[2] R.J. Choudhary,[2] A. Gome,[2] V.R. Reddy,[2] G.R. Umapathy,[3] S. Ojha,[3] E. Carleschi,[4] B.P. Doyle,[4] A.R.E. Prinsloo[1] and C.J. Sheppard[1]

[1]*Cr Research Group, Department of Physics, University of Johannesburg, Johannesburg, PO Box 524, Auckland Park, South Africa*

[2]*UGC-DAE Consortium for Scientific Research, University Campus, Khandwa Road, Indore, MP-452001, India*

[3]*Inter University Accelerator Centre, Aruna Asaf Ali Marg, New Delhi-110067, India*

[4]*ARPES Group, Department of Physics, University of Johannesburg, Johannesburg, PO Box 524, Auckland Park, South Africa*

Cubic spinel $CoCr_2O_4$ has attained recent attention due to its multiferroic properties. However, the Co site substitution effect on the structural and magnetic properties has rarely been studied in thin film form. In this work, the structural and magnetic properties of $Co_{1-x}Ni_xCr_2O_4$ ($x$ = 0, 0.5) epitaxial thin films deposited on $MgAl_2O_4$ (100) and MgO (100) substrates to manipulate the nature of strain in the films using pulsed laser deposition (PLD) technique are presented. The epitaxial nature of the films was confirmed through X-ray diffraction (XRD) and Rutherford backscattering spectrometry (RBS) measurements. Raman measurements revealed a disappearance of characteristic $A_{1g}$ and $F_{2g}$ modes of the $CoCr_2O_4$ with increase in the Ni content. Atomic force microscopy (AFM) studies show a modification of the surface morphology upon Ni substitution. Magnetic measurements disclose that the ferrimagnetic Curie temperature ($T_c$) of the $CoCr_2O_4$ in thin film grown on $MgAl_2O_4$ (100) and MgO (100) substrates were found to be 100.6 ± 0.5 K and 93.8 ± 0.2 K, respectively. With Ni substitution the transition temperatures significantly get enhanced from that of $CoCr_2O_4$. X-ray photoelectron spectroscopy (XPS) suggests $Cr^{3+}$ oxidation states in the films, while Co ions are present in a mixed $Co^{2+}/Co^{3+}$ oxidation state. The substitution of Ni at Co site significantly modifies the line shape of the core level as well as the valence band. Ni ions are also found to be in a mixed 2+/3+ oxidation state. O 1*s* core level display asymmetry related to possible defects like oxygen vacancies in the films.

*Keywords: Chromites, Pulsed laser deposition (PLD), Thin Films, Raman, XPS.*

[*] Corresponding author: pankajm@uj.ac.za

# I. INTRODUCTION

$CoCr_2O_4$ is a ferrimagnetic spinel compound with complex magnetic interactions at low temperatures [1]. The magnetic cations $Co^{2+}$ and $Cr^{3+}$ ions occupy the tetrahedral *A* and octahedral *B* sites, respectively, of the spinel structure [1]. Below Curie temperature ($T_C$ = 93 K), a collinear ferrimagnetic ordering occurs, with further lowering the temperature to 26 K ($T_S$) the compound develops a short-range-ordered (SRO) spiral component [1]. Theoretical calculations has also evidenced the $T_S$ feature due to the nesting of the Fermi surface having high degree of magnetic instability in the system associated with the nesting vector $q$ along the [110] direction [2]. Recent experiments have shown the size dependence of these magnetic transitions [3]. Below a critical particle size of $d_{c,spiral}$ = 4.4 (1) nm, the spiral ordering vanishes [3] and for size less than $d_{c,col}$ = 3.3 (1) nm, the collinear magnetic order disappears [3]. A lock-in transition at a temperature $T_L$ = 15 K appears because of the incommensurate-to-commensurate magnetic phase transition [1]. At $T_L$ the period of the spin spiral eventually 'locks in' to the lattice parameter [1]. This spiral component is responsible for inducing the multiferroicity in $CoCr_2O_4$ [1]. Structurally, $CoCr_2O_4$ is stable and remains cubic with space group $Fd\bar{3}m$ at all temperatures as the sample is cooled down to 10 K [4].

On the other hand for $NiCr_2O_4$ the crystal structure of the system reduces from cubic to tetragonal at 310 K [5, 6]. In $NiCr_2O_4$ the cubic-to-tetragonal phase transition is of first order in nature [7]. It is driven by a cooperative Jahn-Teller (J-T) effect in the array of $Ni^{2+}$ cations in tetrahedral position [7]. For the $Cr^{3+}$ $(3d^3)$ cations the J-T effect in the octahedral crystal field is absent [7]. The tetragonal structure further transforms to an orthorhombic phase, driven by a magneto-structural coupling at the Néel temperature ($T_N$) at about 65 K [8, 9]. Both magnetic susceptibility and heat capacity measurements indicate anomalies at $T$ = 30 K resulting from the distortion occurring within the orthorhombic structure [10].

Investigations into the structural properties of $CoCr_2O_4$ thin films deposited on $MgAl_2O_4$ (001) substrates reveal a self-organized three dimensional isolated pyramidal architecture and hut-clusters with {111} facets [11]. The anisotropy of the surface energy in the spinel is the driving force for the growth of the {111} faceted structure [11]. Aqeel *et al.* [12] reported spin-Hall magnetoresistance (SMR) and spin-Seebeck effect (SSE) in multiferroic $CoCr_2O_4$ thin films with Pt contacts indicating a large enhancement of both the signals occurring below the $T_S$ and $T_L$ transitions. These results indicated that the spin transport at the Pt/$CoCr_2O_4$ interface is sensitive to the magnetic phases, but this cannot be explained solely by the magnetic properties observed in the bulk samples [12]. The magnetic anisotropy in the spinel structure of $CoCr_2O_4$ thin films demonstrate a strain dependence which was supported by theoretical calculations predicting a tensile strain that favors the spin orientation in the film plane whereas a compressive strain induces an out-of-plane magnetic easy axis which is in contradiction with what is reported for $CoFe_2O_4$ [13].

Guzman *et al.* [14] found that the deposition of $CoCr_2O_4$ thin films on $MgAl_2O_4$ (100) and MgO (100) substrates show optimum growth, with good quality films on the spinel $MgAl_2O_4$ substrate, despite the large lattice mismatch of about 3%. On the contrary, $CoCr_2O_4$ thin films deposited on MgO (100) substrate, with a rock-salt structure with lower lattice mismatch (~ 1%), degrades the crystal quality of the pristine films by forming antiphase boundaries (APBs). The type and number of APBs can be manipulated by changing the growth temperature that will ultimately help to partially recover the magnitude of magnetization [14]. By using soft x-ray techniques Windsor *et al.* [15] have shown that the magnetic behavior of a strained [110] oriented $CoCr_2O_4$ film is a type-II multiferroic. The resonant soft x-ray diffraction (RXD) signal of the ($q$, $q$, 0) reflection appeared below $T_S$, the same ordering temperature as the conical magnetic structure in the bulk, indicating that this phase remains multiferroic even under strain [15]. However, reports on doped $CoCr_2O_4$ thin films are lacking. Recently, magnetic and electronic properties of $(Ni_{1-x}Co_x)Cr_2O_4$, with $0 \leq x \leq 1$, were reported [16, 17]. Substituting Co at the Ni site effectively changed the magnetic transition temperatures [16, 17]. In case of nanoparticles of $(Co_{1-x}Ni_x)Cr_2O_4$, with $x = 0.5$ and 0.25, showed modified magnetic transition temperatures along with wasp-waist like features observed in the magnetization as a function of applied magnetic field measurements at low temperatures [18]. In addition, temperature dependent infrared (IR) studies revealed anomalous behavior of phonons below $T_C$ for $Co_{0.9}Ni_{0.1}Cr_2O_4$, attributed to spin-phonon coupling [19].

Having similar molecular formula, both $CoCr_2O_4$ and $NiCr_2O_4$ differs significantly for their distinct properties. Beside ferrimagnetic $T_C$, $CoCr_2O_4$ has different magnetic transitions at low temperatures without any change in its cubic crystal structure. On the other hand, $NiCr_2O_4$ has temperature dependent structural as well as magnetic phase transitions significantly different from that of $CoCr_2O_4$. $NiCr_2O_4$ demonstrates tetragonal crystal structure at room temperature whereas $CoCr_2O_4$ has cubic structure. At present, there is no report on the structural and magnetic properties of Ni doped $CoCr_2O_4$ thin films. While depositing thin films, the role of selection of substrate to manipulate the type of the strain in the film has also been explored in this work. In the present paper, the structural and magnetic phase transitions in $Co_{1-x}Ni_xCr_2O_4$ (with $x = 0, 0.5$) thin films on $MgAl_2O_4$ (100) and MgO (100) single crystal substrates are discussed.

## II. EXPERIMENTAL DETAILS

$Co_{1-x}Ni_xCr_2O_4$, with $x = 0, 0.5$, powder samples were prepared using the sol-gel technique, following the procedure reported earlier [18]. The amorphous powders were then calcined at 900 °C. The $Co_{1-x}Ni_xCr_2O_4$ ($x = 0, 0.5$) targets for pulsed laser deposition (PLD) were then prepared by pressing calcined powder samples into a pellet and sintering these at 1100 °C for 12 h in a programmable tubular furnace. The thin films were deposited from these targets using PLD. After loading the ceramic target and substrates into the PLD chamber, the chamber was evacuated to obtain a base pressure about $10^{-5}$ Torr prior to the deposition. The ceramic targets were ablated using a *KrF* excimer laser (Lambda Physik COMPex 201 Model, Germany) at constant laser energy of 220 mJ and 10 Hz repetition rate. The target to substrate distance was kept constant at four centimeters. Commercial (100) oriented $MgAl_2O_4$ and MgO single crystal substrates with dimensions 10×10×0.5 mm were used. The samples were deposited for 30 minutes at 100 mTorr oxygen partial pressure to compensate the oxygen loss from the targets during the ablation process and to make the films stoichiometric. During deposition, the substrate temperature was maintained at 750 °C and the target was rotated continuously during the laser ablation process. After deposition the films were cooled down to room temperature under the same oxygen partial pressure in order to avoid the formation of cracks.

Structural characterizations of these samples were carried out by utilizing high resolution X-ray diffraction (HRXRD) techniques, with a Bruker D 8 X-ray diffractometer and using a Cu–$K_\alpha$ radiation ($\lambda$ = 1.54056 Å). Raman measurements (JOBIN YVON HORIBA HR 800) were performed using a diode laser having wavelength $\lambda$ = 473 nm (Power 25 mW, grating 1800). Atomic force microscopy (AFM) measurements were done using a Multimode Atomic Force Microscope employing a $Si_3N_4$ tip. The magnetic measurements were performed using a SQUID-VSM from Quantum Design (USA). The compositions of the films were determined by the Rutherford backscattering spectrometry (RBS) measurements using α-particles ($He^{2+}$) of energy 2 MeV and the accumulated charge on the film was 20 $\mu$C. X-ray photoemission spectroscopy (XPS) measurements were acquired at room temperature using a SPECS XR 50M monochromatised X-ray source equipped with an Al $K_\alpha$ anode ($h\nu$ = 1486.71 eV) and a SPECS PHOIBOS 150 hemispherical electron energy analyser. The base pressure of the experimental chamber was < 2 × $10^{-10}$ mbar. Surface charge compensation was obtained by using a low-energy electron flood gun (electron energy = 3 eV, electron flux = 20 $\mu$A). The overall energy resolution of the XPS spectra was set to 0.7 eV for the survey scans, and to 0.55 eV for all the other spectra presented in this work.

## III. RESULTS AND DISCUSSION

The motivation for using two different types of substrates such as MgO and $MgAl_2O_4$ for the deposition of films was to induce two different types of strain in the films [13]. In case, the lattice parameter of the film matches exactly with the lattice parameter of the substrate, then epitaxial growth of the film occurs as shown in Fig. 1 (a) considering a simple cubic substrate and the film. If there is a slight mismatch of the lattice parameters between the film and the substrate then two cases arise. Case (i): Film grows on the substrate retaining the lattice mismatch (Fig. 1 (b): $a_{\mathrm{sub}} < a_{\mathrm{film}}$ and (d): $a_{\mathrm{sub}} > a_{\mathrm{film}}$) and case (ii): The lateral lattice parameter of the film distorts to minimize the strain energy and matches the lattice parameter of the substrate (Fig. 1 (d) and (e)). The situation as in case (i) where the film retains the bulk lattice parameter is termed as “relaxed films” and if the film distorts to match the lattice parameter of the film as indicated in case (ii) are designated as “strained films”. Depending on the type of lattice mismatch consequent strain can be of two types such as compressive and tensile. The film having lattice parameter higher than the substrate as shown in Fig. 1 (b) if forced to match the lattice parameter of the substrate (Fig. 1 (d)), the nature of strain would be compressive. On the other hand the film having lattice parameter smaller than the substrate as indicated in Fig. 1 (c) if compelled to match the lattice parameter of the substrate (Fig. 1 (e)), the type of strain in the film will be termed as tensile strain. $MgAl_2O_4$ has a spinel structure analogue to that of $CoCr_2O_4$ but with a smaller lattice parameter that causes a lattice mismatch of 3.1% that induces a large in-plane compressive strain in the films [13] similar to the case as shown in Fig. 1 (d). On the other hand, MgO (100) has a rock-salt crystal structure that has a lattice mismatch of 1.1% with $CoCr_2O_4$ which will induce a tensile strain in the films [13] analogues to the situation as depicted in Fig. 1 (e). Despite having a significant lattice mismatch between both the substrates and the $Co_{1-x}Ni_xCr_2O_4$ (with $x$ = 0, 0.5), that can create strain, good quality thin films have been grown on both $MgAl_2O_4$ (100) and MgO (100) substrates, as will be discussed later. The effect of inducing strain in the spinel material by depositing on substrates having lattice mismatch was also observed previously in the case of $CoFe_2O_4$ [20]. $CoCr_2O_4$ thin films were previously grown on [001]-oriented $MgAl_2O_4$ and MgO substrates by Heuver *et al.* [13]. Studies revealed the structure and magnetism of these films with theoretical estimation of strain to support their results. However, the effect of Ni substitution on the crystal structure, magnetic and electronic properties is not reported.

The XRD patterns of the $Co_{1-x}Ni_xCr_2O_4$ ($x$ = 0, 0.5) films deposited on $MgAl_2O_4$ (100) substrates are shown in Fig. 2 (a) and MgO (100) substrates are shown in Fig. 2 (b). It indicates a reflection corresponding to the (400) plane of the $CoCr_2O_4$ on the two different substrates. The presence of only (400) reflection peaks suggest epitaxial nature which is also confirmed from the RSM data as discussed below, of the film in line with the findings of Heuver *et al.* [13]. The XRD pattern of both MgO (100) and $MgAl_2O_4$ (100) substrates are also shown in order to visualize the peak related to these substrates. The splitting seen in the most intense peaks of both the substrates are attributed to the Cu–$K_{\alpha1}$ and Cu–$K_{\alpha2}$ split [21]. The out-of-plane lattice constants of the films are given in Table I, where the bulk value of $c$ is 8.33 Å [13]. The lattice parameter $c$, for both the compositions, has a smaller value (Table I) for the films deposited on the MgO (100) substrate when compared to that obtained for the films grown on $MgAl_2O_4$ (100). This is in line with the findings of Heuver *et al.* [13]. Upon Ni substitution the (400) reflection shifts toward higher 2θ values for both the substrates resulting decrease in lattice parameter (see Fig. 2). As Ni has higher atomic radius compared to Co, the shifting of peak towards higher diffraction angle resulting decrease in lattice parameters corroborates the incorporation of Ni at Co site.

In order to calculate the thickness of the films X-ray reflectivity (XRR) measurements were done (shown in Fig. 3). The oscillations observed in the XRR caused due to the interference of the X-ray beam reflected from the surface of the film and from the film and substrate interface. From the oscillations, in the reflectivity curve the thickness of the films was calculated and these values are given in Table II. Fig. 3 (d) depicts the fitting of the observed data for the $(Co_{0.5}Ni_{0.5})Cr_2O_4$ thin film grown on MgO and $MgAl_2O_4$ substrates using Parratt's formalism [22]. The blue line in the figure denotes the simulated curve.

Reciprocal space mapping (RSM) measurement is used to exactly determine properties, such as the strain state and the in-plane lattice parameters of the thin films. This is performed by analyzing the distribution of the diffraction intensity near a reciprocal lattice point of a substrate and film [23, 24]. The RSM data measured around asymmetric (115) lattice point indicates that all the films are epitaxial in nature. One can get an idea about the strain state of the film from such an asymmetric scan and it is to be noted that if the film is fully strained then one would expect the lattice point corresponding to film to occur at the same $q_x$ as that of substrate. This is shown by a dashed line in Fig. 4. The film reciprocal lattice point move away from this same $q_x$ line as film relaxes and for fully relaxed film then the film reciprocal lattice point lies along the line of relaxation. The data of the present work suggests that all the films are partially strained, however the $(Co_{0.5}Ni_{0.5})Cr_2O_4$ film grown on $MgAl_2O_4$ (100) substrate exhibiting more strain as compared to the remaining samples.

As $CoCr_2O_4$ crystallizes in $Fd\bar{3}m$ crystal structure, Raman modes are predicted by group theory for this space group [25]. Group theoretical calculations predict 17 fundamental lattice vibration modes [25]. The representations of these modes are expressed as [25]:

$$\Gamma = A_{1g} + E_g + F_{1g} + 3F_{2g} + 2\,A_{2u} + 2E_u + 5F_{1u} + 2F_{2u} \quad \ldots\ldots\ldots\ldots.(1)$$

Among these modes, $A_{1g}$, $E_g$, $3F_{2g}$ modes are Raman active whereas $4F_{1u}$ modes are IR-active [25]. Experimentally, only five Raman modes at 692 $cm^{-1}$ ($A_{1g}$), at 457 $cm^{-1}$ ($E_g$), and at 610, 515 and at 186 $cm^{-1}$ ($F_{2g}$) [26] were observed for $ZnCr_2O_4$ single crystal analogous to the crystal structure of $CoCr_2O_4$ [26].

In Fig. 5 (a) and (b) the Raman spectra are shown for $CoCr_2O_4$ and $Co_{0.5}Ni_{0.5}Cr_2O_4$ films deposited on $MgAl_2O_4$ (100) and MgO (100) substrates, respectively. The Raman modes related to the substrate are observed at 409 $cm^{-1}$, 667 $cm^{-1}$ 720 $cm^{-1}$, and 766 $cm^{-1}$ for $MgAl_2O_4$ (100), while no modes are observed for the MgO (100) substrate. No noticeable Raman modes up to 900 $cm^{-1}$ wave number range and relatively low back ground intensity makes MgO one of the most suitable substrate in this prospect [27, 28]. For $CoCr_2O_4$ deposited on both substrates, significant Raman modes are observed at 551 $cm^{-1}$ related to $F_{2g}$ mode. The peak observed at 551 $cm^{-1}$ is shifted by 35 $cm^{-1}$. This shift to higher wave numbers in the $F_{2g}$ mode has also been observed in $CoCr_2O_4$ nanoparticles [29] and is attributed to the size effect [29]. The observed Raman modes were labeled in Fig. 5. For $Co_{0.5}Ni_{0.5}Cr_2O_4$ films deposited on both substrates no Raman modes are observed. As phonons are sensitive to the finite length scale effects due to the long range order that can be influenced by the strain, cation redistribution, phonon confinement effects, directional nature and other factors [29, 30], with the substitution of Co by Ni in $CoCr_2O_4$ the Raman modes disappears possibly indicating structural disorder that sets in by the substitution [31].

In order to investigate the surface morphology of the thin films, AFM measurements were carried out on the thin films. These are shown in Fig. 6. Considering Figs. 6 (a) and (b) for the $CoCr_2O_4$ thin films grown on $MgAl_2O_4$ (100) and MgO (100) substrates shows similar morphology with distinct grains. Upon Ni substitution, the quality of the surface degraded resulting quite hazy AFM image as can be seen in Figs. 6 (c) and (d). The average roughness of the $CoCr_2O_4$ thin films grown on $MgAl_2O_4$ (100) and MgO (100) substrate ranges from 0.8 to 1.2 nm. Upon Ni substituting the average roughness was found to decrease to values ranging from 0.5 to 0.7 nm. However, upon careful observation, it can be seen that the grain size has remarkably increased with Ni substitution (Fig. 6).

In order to explore the elemental distribution, thickness and epitaxial nature of the deposited thin films Rutherford backscattering spectrometry (RBS) was performed under both random and channeling conditions. RBS measurement of an Au film was taken for the calibration before the sample measurements were started. RBS measurement involves bombardment of high energy $He^{2+}$ ions known as $\alpha$-particles on a sample and the energy and yield of back scattered $\alpha$-particles are noted. From the determination of the energy of the back-scattered particles the mass of the atoms from which these α-particles were scattered and the depth at which such scattering occurred can be traced. The concentration of particular element is directly proportional to the number of scattered $\alpha$-particles. The RBS results are depicted in Fig. 6. The desired elements present in the thin films, as well as in the substrate, were identified in the spectra, where the width of each peak corresponds to the relative thickness of that material in the film. RBS techniques are popularly used for estimation of the chemical composition, thickness and atomic distribution of the elements throughout the film thickness and composition [32]. If the material or film is oriented along the substrate, and the direction of $\alpha$-particles was aligned in the substrate orientation direction then most of the alpha particles will pass through the channel or empty space available within the crystal lattice leading to lower yield [33]. The fundamental principle is that the energetic ions can move inside the crystals that may penetrate much deeper if the atoms in the crystal are directed in some specific crystal directions [34]. This phenomenon is termed as a "channeling" effect that can be utilized for instance in specific ion beam analysis methods and can be explained analytically and atomistic simulations based on molecular dynamics (MD) methods [34]. From Fig. 7 it is clear that the yield has decreased significantly when the film is aligned and channeling occurs. However, this kind of decrease in yield cannot be observed for a polycrystalline thin film which has not grown along a preferred orientation [35]. The presence of channeling supports the XRD and RSM findings and supports the epitaxial growth of the films. The peak profile of Ni doped $CoCr_2O_4$ on MgO substrates showed diffused nature indicating non uniform distribution of both Ni and Co throughout the film thickness (see Fig. 7 (d)). The thicknesses of the films were calculated by fitting the RBS random spectra using the RUMP simulation code [36]. The values obtained are tabulated in Table III. Except for the estimated thickness obtained for the $CoCr_2O_4$ grown on MgO (100) substrate, all other values were almost double the value predicted from the XRR measurements. These discrepancies arise because of the different principles that are involved in XRR and RBS. In literature, often these techniques yield different results for the same film [37, 38]. For instance, the thickness of $HfO_2$ film was determined to be 0.21 nm from RBS, whereas XRR showed the thickness to be 1.2 nm [37]. However, it is believed that XRR can be considered to evaluate the thickness more precisely when compared to the RBS technique due to its high-density contrast [37]. The typical simulated curves for $CoCr_2O_4$ thin film grown on two substrates are also depicted in Fig. 8 (see red curves). Form Fig. 8 it is also evident that the presence of Co is bit non-uniform in the films and has higher concentration at the film surface. For the doped samples, the fitting became quite difficult due to complex nature of Co and Ni distribution in the film, with the Co and the Ni having nearly similar atomic masses. The yield from the film is reduced because the channeling was performed along the substrate, thereby confirming the epitaxial nature of the films.

XPS measurements were performed in order to investigate the electronic structure and the oxidation states of the different chemical elements in the thin films. Fig. 9 shows the wide survey spectra for the four investigated samples. Spectra show all the expected peaks and no additional elements were detected. Photoemission and Auger peaks are labeled accordingly in the figure. Co and Ni core levels vary in intensity according to the sample composition, as expected.

Fig. 10 shows the O 1*s* XPS spectra for the investigated thin films. Each spectrum has been fitted with three Voigt-type single components - labeled as O(1), O(2) and O(3) as shown in Fig. 10 - in addition to a Shirley background. This is consistent with what published in the literature for similar systems [39, 40]. Component O(1), on the lower binding energy side of the spectrum, is attributed to stoichiometric oxygen in the oxide main matrix. The second component O(2) is ascribed to oxygen vacancies or defects in the lattice [39, 40], while O(3) can be ascribed to residual chemisorbed oxygen [41]. The fitted binding energy (BE) and the relative percentage area of each component are reported in Table IV, and they show consistent values for the four investigated samples.

Fitted Cr 2*p* core level spectra are shown in Fig. 11. The Cr 2*p* BE region is composed of two main asymmetric peaks located at ~576 eV and ~586 eV, corresponding to the Cr $2p_{3/2}$ and Cr $2p_{1/2}$ spin orbit components respectively. The line shape of the Cr 2*p* spectrum does not show significant changes for the four samples. Each spectrum was fitted with three spin-orbit doublets - labeled as Cr(1), Cr(2) and Cr(3) - a Shirley-type background, a broad satellite structure at about 597 eV BE. These, together with the overall fit to the data, are all reported in Fig. 11. This fine multiplet structure, composed of three multiplet lines, is consistent with Cr ions being in a 3+ oxidation state [42, 43] as previously reported for many Cr-based oxides. The fitted binding energy (BE), the spin orbit energy splitting ($\Delta_{SO}$) and the relative percentage area for Cr(1), Cr(2) and Cr(3) are reported in Table V.

Fig. 12 shows the fitted Ni 2*p* core level spectra for $(Co_{0.5}Ni_{0.5})Cr_2O_4$ grown on (a) $MgAl_2O_4$ (100) and (b) MgO (100). The line shape of this core level is composed of two main peaks with centroids located at about 855 eV and 873 eV, corresponding to Ni $2p_{3/2}$ and Cr $2p_{1/2}$ spin orbit components respectively. Together with these, two broad satellite features are located at about 7 eV higher BE from the main peaks. The overall fit to the experimental data for both spectra was obtained by using two main spin orbit doublets, two satellite doublets, and a Shirley-type background. The two main doublets correspond to Ni ions in 2+ and 3+ oxidation states, and have therefore been labeled as $Ni^{2+}$ and $Ni^{3+}$ in Fig. 12. The fit parameters for this core level are reported in Table VI. It is important to note that the area ratio $Ni^{3+}$:$Ni^{2+}$ is 2:1 for both thin films. Interestingly, this ratio is slightly smaller than the corresponding component in nanoparticle form [18].

A comparison of Co 2*p* core level spectra is reported in Fig. 13. The line shape of this core level is composed of two main peaks at ~780 eV and ~795 eV, corresponding to Co $2p_{3/2}$ and Co $2p_{1/2}$ spin-orbit components; each component has a higher BE broad charge transfer satellite, which extends from 5 eV to 10 eV higher BE from the main peak. For both $CoCr_2O_4$ thin films – spectra (a) and (c) in Fig. 13 – the line shape of the Co 2*p* core level resembles that of $Co_3O_4$ [44, 45]. The intensity maxima for $2p_{3/2}$ and $2p_{1/2}$ are located at 780 eV and 795 eV respectively, and the main peaks show an asymmetry on the higher BE side. The $2p_{3/2}$ satellites show a double feature with centroids at 785.7 eV and 789.3 eV; the double feature in the $2p_{1/2}$ satellite is less evident but still present. This is a fingerprint of a mixed $Co^{2+}$/$Co^{3+}$ oxidation state for Co ions in these samples. The introduction of 50% Ni doping in replacement of Co ions modifies quite drastically the line shape of this core level (as opposed to Cr 2*p* as stated above), see spectra (b) and (d) in Fig. 13. The centroid of the main peaks shifts to 0.5 eV higher BE. The main peaks become broader and more symmetric, and the charge transfer satellites appear more intense compared to the main peaks. Furthermore, the difference in energy between the main peaks and their satellites is now ~5.3 eV, as opposed to 6.2 eV in spectra (a) and (c). Spectra (b) and (d) resemble more the line shape of the typical charge-transfer-type compound CoO [46]. This suggests that the presence of Ni has induced a pronounced change in the local electronic structure of the Ni-doped thin films, favoring a more localized for Co 3*d* electrons (seen in the increase in the intensity of the charge-transfer satellite in Co 2*p* core level), and possibly an increase in the presence of Co ions in a 2+ oxidation state.

Finally, Fig. 14 shows a comparison of the valence band region for the four investigated thin films. The line shape of the valence band for both $CoCr_2O_4$ thin films consists of a sharp feature located between the Fermi level ($E_F$, BE = 0 eV) and 2 eV BE, which can be mostly ascribed to a superposition of Cr 3*d* and Co 3*d* partial density of states [47-49]. This peak is separated from a wider structured region between ~2 eV and ~8 eV, where O 2*p* states are expected to be located [48]. The broad region between ~8 eV and ~12 eV is likely to derive from Co/Cr 4s states. The partial substitution of Co by Ni ions in both $(Co_{0.5}Ni_{0.5})Cr_2O_4$ thin films have the effect of (1) slightly depleting the spectral weight of the above-mentioned near-$E_F$ feature (due to the decreased contribution and increased localization of Co 3*d* states, in agreement with the analysis of the Co 2*p* core level spectra reported above), and (2) increasing the spectral weight in the BE region from 1 eV to 3 eV, where the Ni 3*d* states are generally located [50]. Although the valence bands are measured 300 K, the significant change in the line shape upon Ni substitution is somehow related to enhancement of the $T_C$ for the $(Co_{0.5}Ni_{0.5})Cr_2O_4$ thin films grown on $MgAl_2O_4$ and MgO substrates. To further explore the magnetic properties, in-plane magnetic measurements were carried out using a SQUID-VSM.

Temperature dependent in-plane magnetization measurements, shown in Fig. 15, were performed with a probing magnetic field of 100 Oe. Films deposited on $MgAl_2O_4$ substrates, having compressive strain, show a ferrimagnetic $T_C$ = 100.6 ± 0.5 K; whereas the film grown under tensile strain (grown on MgO (100)) show $T_C$ = 93.8 ± 0.2 K for $CoCr_2O_4$ (Fig. 15). The bulk $T_C$ value for $CoCr_2O_4$ is 93 K [1]. The transition is prominent for the $CoCr_2O_4$ film grown on MgO (100) substrate. In order to obtain $T_C$, d$M$/d$T$ as function of $T$ plots were used and the $T_C$ values were taken as the temperature associated with the minimum (insets of Fig. 15 (a) and (c)). The values of $T_C$ were found to be 88.4 K for the film deposited on the $MgAl_2O_4$ (100) substrate and 91.9 K for the film grown on the MgO (100) substrate. However, upon substituting Ni at Co site, the transitions appeared to be smeared out (see Fig. 14 (a) and (b)). The $T_C$ values were determined to be 104.5 ± 0.4 K and 108.5 ± 0.6 K for films grown on the $MgAl_2O_4$ (100) and MgO (100) substrates, respectively, (see insets of Fig. 15 (b) and (d)) using knee point method as described previously [16, 17]. Comparing $T_C$ values found for the films compared with the powder samples previously measured are given in Table VII. These were determined using the knee-point method as reported previously for all the samples [16, 17]. The reported $T_C$ values for $CoCr_2O_4$ films grown on $MgAl_2O_4$ (100) and MgO (100) substrates were previously [13] found to be ~ 81 K and ~ 80 K, respectively, considering the 0.01 T probing field along [100] in plane direction. The magnitude of magnetic moment of the film is small due to the thin layer of the film, as can be seen from the order of magnitude of magnetic moments (Fig. 15). Magnetic properties are greatly influenced by the cationic site occupancy and anisotropy induced by the substrate [13]. For the film grown on the MgO (100) substrate, where the film experience tensile strain, the $T_C$ value was found to be 93.8 ± 0.2 K for $CoCr_2O_4$. This is less when compared to the bulk value. However, for the film grown on the $MgAl_2O_4$ (100) substrate, experiencing compressive stress, the $T_C$ value was enhanced to 100.6 ± 0.2 K. This difference in $T_C$'s for different substrates is significant compared to the earlier reports [13]. It is interesting to note that upon Ni substitution causes the reverse in magnitude of the $T_C$ when the substrates are changed. This clearly indicates the magnetic properties are controlled by the anisotropy in addition to the dopants [16]. Further, to explore the magnetism below $T_C$, $M$ versus $\mu_0 H$ measurements were performed (Fig. 16). Fig. 16 (a) shows the hysteresis loop measured at 20 K for the $COCr_2O_4$ film grown on MgO (100) substrate. The constricted loop around the low field region is possibly due to the spiral ordering of the spins [1, 2]. When the measurement temperature is increased to 60 K (Fig. 16 (b)), the irreversibility of the loop still persists indicating the ferrimagnetic ordering as observed in the bulk [1, 2]. The diamagnetic tailing of the hysteresis loop is due to the contribution from the substrate. As the magnetic moment of the film is quite less, the substrate constriction dominates. The hysteresis loop measured for the $CoCr_2O_4$ thin film grown on $MgAl_2O_4$ substrate at 60 K also shows irreversibility (Fig. 16 (c)). Upon Ni substitution at Co site retains the ferrimagnetic order below $T_C$ as shown in Fig. 16 (d) and (e) measured at 60 K.

## IV. CONCLUSIONS

Thin films of $Co_{1-x}Ni_xCr_2O_4$ ($x$ = 0, 0.5) were deposited on $MgAl_2O_4$ (100) and MgO (100) substrates using PLD technique. The films demonstrate phase purity without any contamination. With Ni substitution, characteristic Raman modes disappear indicating disorder or possible structural changes. XRD and RBS measurements confirm oriented epitaxial nature of the thin films. The observation of oscillations in the X-ray reflectivity of the films suggest the deposition of the thin films that causes the interference fringes. AFM studies reveal the change of surface morphology upon Ni substitution. XPS studies demonstrate the complex nature of the oxidation states of Co and Ni. Magnetic measurements reveal the ferrimagnetic transition temperature, $T_C$, of the $CoCr_2O_4$ is modified when grown on different substrates that have induced different types of anisotropies into the films. In addition substitution of Co site with Ni has found to change the $T_C$ of the films which is also evidenced from the change in the line shape of the valence band.

**ACKNOWLEDGEMENTS**

This work was supported by South African National Research Foundation (NRF Grant No's: 80880, 93205, 85365, 90698 and 93205) and the URC and FRC of UJ, South Africa. NRF, South Africa is acknowledged for the travel support (Grant No: 109896) to carry out the experiments at UGC-DAE, Consortium for Scientific Research, Indore, India. Authors thank Dr. V.G. Sathe, UGC-DAE, Consortium for Scientific Research, Indore, India for the Raman measurements.

---

**TABLE CAPTIONS**

**TABLE I.** Lattice parameters of the thin film samples deposited on various substrates as calculated from XRD results.

**TABLE II.** Thickness of the films as calculated from the XRR and RBS measurements.

**TABLE III.** Thickness of the films as calculated from RBS results.

**TABLE IV.** Fit results for O 1*s* core level.

**TABLE V.** Fit results for Cr 2*p* core level.

**TABLE VI.** Fit results for Ni 2*p* core level.

**TABLE VII.** Curie temperatures ($T_C$) determined from $M(T)$ measurements for the various samples.

**TABLE I.** Lattice parameters of the thin film samples deposited on various substrates as calculated from XRD results.

| Sample | Lattice parameter ($a$) determined for the films deposited on $MgAl_2O_4$ (100) with $a$ = 8.08 Å | Lattice parameter ($a$) determined for the films deposited on MgO (100) $a$ = 4.21 Å | Remarks |
|---|---|---|---|
| $CoCr_2O_4$ | 8.36 Å | 8.25 Å | This work |
| $(Co_{0.5}Ni_{0.5})Cr_2O_4$ | 8.24 Å | 8.21 Å | This work |
| $CoCr_2O_4$ | 8.58 Å | 8.17Å | Heuver *et al.* [13] |

**TABLE II.** Thickness of films as calculated from the XRR measurements.

| Sample | Thickness of films grown on $MgAl_2O_4$ (100) | Thickness of films grown on MgO (100) |
|---|---|---|
| $CoCr_2O_4$ | ~ 394 Å | ~ 666 Å |
| $(Co_{0.5}Ni_{0.5})Cr_2O_4$ | ~ 235 Å | ~ 344 Å |

**TABLE III.** Thickness of films as calculated from RBS results.

| Sample | Thickness of films grown on $MgAl_2O_4$ (100) substrate | Thickness of films grown on MgO (100) substrate |
|---|---|---|
| $CoCr_2O_4$ | ~ 70 nm | ~ 70 nm |
| $(Co_{0.5}Ni_{0.5})Cr_2O_4$ | ~ 70 nm | ~ 80 nm |

**TABLE IV.** Fit results for O 1*s* core level.

| Sample | O(1) BE (eV) | O(2) BE (eV) | O(3) BE (eV) | O(1) % area | O(2) % area | O(3) % area |
|---|---|---|---|---|---|---|
| $CoCr_2O_4$ on $MgAl_2O_4$ (100) | 529.82 | 531.51 | 533.20 | 58 | 38 | 4 |
| $(Co_{0.5}Ni_{0.5})Cr_2O_4$ on $MgAl_2O_4$ (100) | 529.88 | 531.79 | 533.21 | 55 | 41 | 4 |
| $CoCr_2O_4$ on MgO (100) | 529.85 | 532.42 | 533.30 | 62 | 35 | 3 |
| $(Co_{0.5}Ni_{0.5})Cr_2O_4$ on MgO (100) | 529.87 | 531.53 | 533.22 | 61 | 36 | 3 |

**TABLE V.** Fit results for Cr 2*p* core level.

| Sample | Cr(1) BE (eV) | Cr(2) BE (eV) | Cr(3) BE (eV) | Cr(1) $\Delta_{SO}$ (eV) | Cr(2) $\Delta_{SO}$ (eV) | Cr(3) $\Delta_{SO}$ (eV) | Cr(1) % area | Cr(2) % area | Cr(3) % area |
|---|---|---|---|---|---|---|---|---|---|
| $CoCr_2O_4$ on $MgAl_2O_4$ (100) | 575.09 | 576.24 | 577.63 | 9.81 | 9.84 | 9.88 | 24 | 49 | 27 |
| $(Co_{0.5}Ni_{0.5})Cr_2O_4$ on $MgAl_2O_4$ (100) | 575.04 | 576.25 | 577.64 | 9.81 | 9.84 | 9.88 | 23 | 49 | 28 |
| $CoCr_2O_4$ on MgO (100) | 575.01 | 576.16 | 577.61 | 9.81 | 9.84 | 9.88 | 27 | 50 | 23 |
| $(Co_{0.5}Ni_{0.5})Cr_2O_4$ on MgO (100) | 574.90 | 576.11 | 577.52 | 9.72 | 9.77 | 9.78 | 23 | 46 | 31 |

**TABLE VI.** Fit results for Ni 2*p* core level.

| Sample | $Ni^{2+}$ $2p_{3/2}$ BE (eV) | $Ni^{3+}$ $2p_{3/2}$ BE (eV) | $Ni^{2+}$ $\Delta_{SO}$ (eV) | $Ni^{3+}$ $\Delta_{SO}$ (eV) | $Ni^{2+}$ % area | $Ni^{3+}$ % area |
|---|---|---|---|---|---|---|
| $(Co_{0.5}Ni_{0.5})Cr_2O_4$ on $MgAl_2O_4$ (100) | 854.45 | 855.91 | 17.40 | 17.50 | 37 | 63 |
| $(Co_{0.5}Ni_{0.5})Cr_2O_4$ on MgO (100) | 854.62 | 856.13 | 17.40 | 17.50 | 35 | 65 |

**TABLE VII.** Curie temperatures ($T_C$) determined from $M(T)$ measurements for the various samples.

| Sample | $T_C$ value for film on $MgAl_2O_4$ (100) substrate | $T_C$ value for film on MgO (100) substrate | $T_C$ values for powder samples [14] |
|---|---|---|---|
| $CoCr_2O_4$ | 100.6 ± 0.5 K | 93.8 ± 0.2 K | 99.5 ± 0.5 K |
| $(Co_{0.5}Ni_{0.5})Cr_2O_4$ | 104.5 ± 0.4 K | 108.5 ± 0.6 K | 89.2 ± 0.7 K |
| $CoCr_2O_4$ | 81 K [Ref. 13] | 80 K [Ref. 13] | |

**Figure Captions:**

FIG. 1. A typical cubic crystal lattice of the film (red) grows on substrate (blue) having lattice parameter: (a) $a_{sub} = a_{film}$, (b) $a_{sub} < a_{film}$ ($MgAl_2O_4$ case) and (d): $a_{sub} > a_{film}$ (MgO case: Here "$a$" is considered to be the twice of actual lattice parameter of MgO). The lateral lattice parameters of the film are equal to the bulk values– the film is said to be relaxed ((b) and (d)). The lattice of the film distorts in lateral direction to minimize the strain energy forced to match the lateral lattice parameter of the substrate ((c) and (e)).

FIG. 2. XRD patterns of the $Co_{1-x}Ni_xCr_2O_4$ ($x$ = 0, 0.5) films grown on (a) $MgAl_2O_4$ (100) and (b) MgO (100) substrates.

FIG. 3. X-ray reflectivity of the $Co_{1-x}Ni_xCr_2O_4$ ($x$ = 0, 0.5) films grown on (a-c) $MgAl_2O_4$ (100) and (b-d) MgO (100) substrates.

FIG. 4. XRD Reciprocal space maps (RSMs) around: (a) (115) and (200) reflections of the $CoCr_2O_4$ film grown on $MgAl_2O_4$ (100), (b) (115) reflection of the $CoCr_2O_4$ film grown on MgO (100), (c) (115) and (200) reflections of the $(Co_{0.5}Ni_{0.5})Cr_2O_4$ film grown on $MgAl_2O_4$ (100) and (d) (115) reflection of the $(Co_{0.5}Ni_{0.5})Cr_2O_4$ film grown on MgO (100).

FIG. 5. Raman modes of $Co_{1-x}Ni_xCr_2O_4$ ($x$ = 0, 0.5) films grown on (a) $MgAl_2O_4$ (100) and (b) MgO (100) substrates.

FIG. 6. AFM images of thin films grown on various substrates: (a) $CoCr_2O_4$ on $MgAl_2O_4$ (100), (b) $CoCr_2O_4$ on MgO (100), (c) $(Co_{0.5}Ni_{0.5})Cr_2O_4$ on $MgAl_2O_4$ (100), (d) $(Co_{0.5}Ni_{0.5})Cr_2O_4$ on MgO (100).

FIG. 7. RBS data for thin films grown on various substrates and measured under random and channeling conditions: (a) $CoCr_2O_4$ on $MgAl_2O_4$ (100), (b) $CoCr_2O_4$ on MgO (100), (c) $(Co_{0.5}Ni_{0.5})Cr_2O_4$ on $MgAl_2O_4$ (100), (d) $(Co_{0.5}Ni_{0.5})Cr_2O_4$ on MgO (100).

FIG. 8. RBS data shown in black and fitted curves obtained using RUMP simulation code for $CoCr_2O_4$ thin film grown on (a) $MgAl_2O_4$ (100) and (b) MgO (100) substrates.

FIG. 9. XPS survey scan for (a) $CoCr_2O_4$ on $MgAl_2O_4$ (100), (b) $(Co_{0.5}Ni_{0.5})Cr_2O_4$ on $MgAl_2O_4$ (100), (c) $CoCr_2O_4$ on MgO (100), and (d) $(Co_{0.5}Ni_{0.5})Cr_2O_4$ on MgO (100).

FIG. 10. O 1$s$ XPS core level spectra for (a) $CoCr_2O_4$ on $MgAl_2O_4$ (100), (b) $(Co_{0.5}Ni_{0.5})Cr_2O_4$ on $MgAl_2O_4$ (100), (c) $CoCr_2O_4$ on MgO (100), and (d) $(Co_{0.5}Ni_{0.5})Cr_2O_4$ on MgO (100).

FIG. 11. Cr 2$p$ XPS core level spectra for (a) $CoCr_2O_4$ on $MgAl_2O_4$ (100), (b) $(Co_{0.5}Ni_{0.5})Cr_2O_4$ on $MgAl_2O_4$ (100), (c) $CoCr_2O_4$ on MgO (100), and (d) $(Co_{0.5}Ni_{0.5})Cr_2O_4$ on MgO (100).

FIG. 12. Ni 2*p* XPS core level spectra for (a) $(Co_{0.5}Ni_{0.5})Cr_2O_4$ on $MgAl_2O_4$ (100), and (b) $(Co_{0.5}Ni_{0.5})Cr_2O_4$ on MgO (100).

FIG. 13. Co 2*p* XPS core level spectra for (a) $CoCr_2O_4$ on $MgAl_2O_4$ (100), (b) $(Co_{0.5}Ni_{0.5})Cr_2O_4$ on $MgAl_2O_4$ (100), (c) $CoCr_2O_4$ on MgO (100), and (d) $(Co_{0.5}Ni_{0.5})Cr_2O_4$ on MgO (100).

FIG. 14. Valence band spectra for (a) $CoCr_2O_4$ and $(Co_{0.5}Ni_{0.5})Cr_2O_4$ on $MgAl_2O_4$ (100), and (b) $CoCr_2O_4$ and $(Co_{0.5}Ni_{0.5})Cr_2O_4$ on MgO (100).

FIG. 15. Magnetization as a function of temperature for thin films grown on various substrates: (a) $CoCr_2O_4$ on $MgAl_2O_4$ (100), (b) $CoCr_2O_4$ on MgO (100), (c) $(Co_{0.5}Ni_{0.5})Cr_2O_4$ on $MgAl_2O_4$ (100), (d) $(Co_{0.5}Ni_{0.5})Cr_2O_4$ on MgO (100). Blue symbols represent the data measured under ZFC and red symbols indicate FC conditions.

FIG. 16. Magnetization as a function of applied magnetic field for thin films grown on various substrates: (a) $CoCr_2O_4$ on MgO (100) (T = 20 K), (b) $CoCr_2O_4$ on MgO (100) (T = 60 K), (c) $CoCr_2O_4$ on $MgAl_2O_4$ (100) (T = 60 K), (d) $(Co_{0.5}Ni_{0.5})Cr_2O_4$ on $MgAl_2O_4$ (100) (T = 60 K) and (e) $(Co_{0.5}Ni_{0.5})Cr_2O_4$ on MgO (100) (T = 60 K).

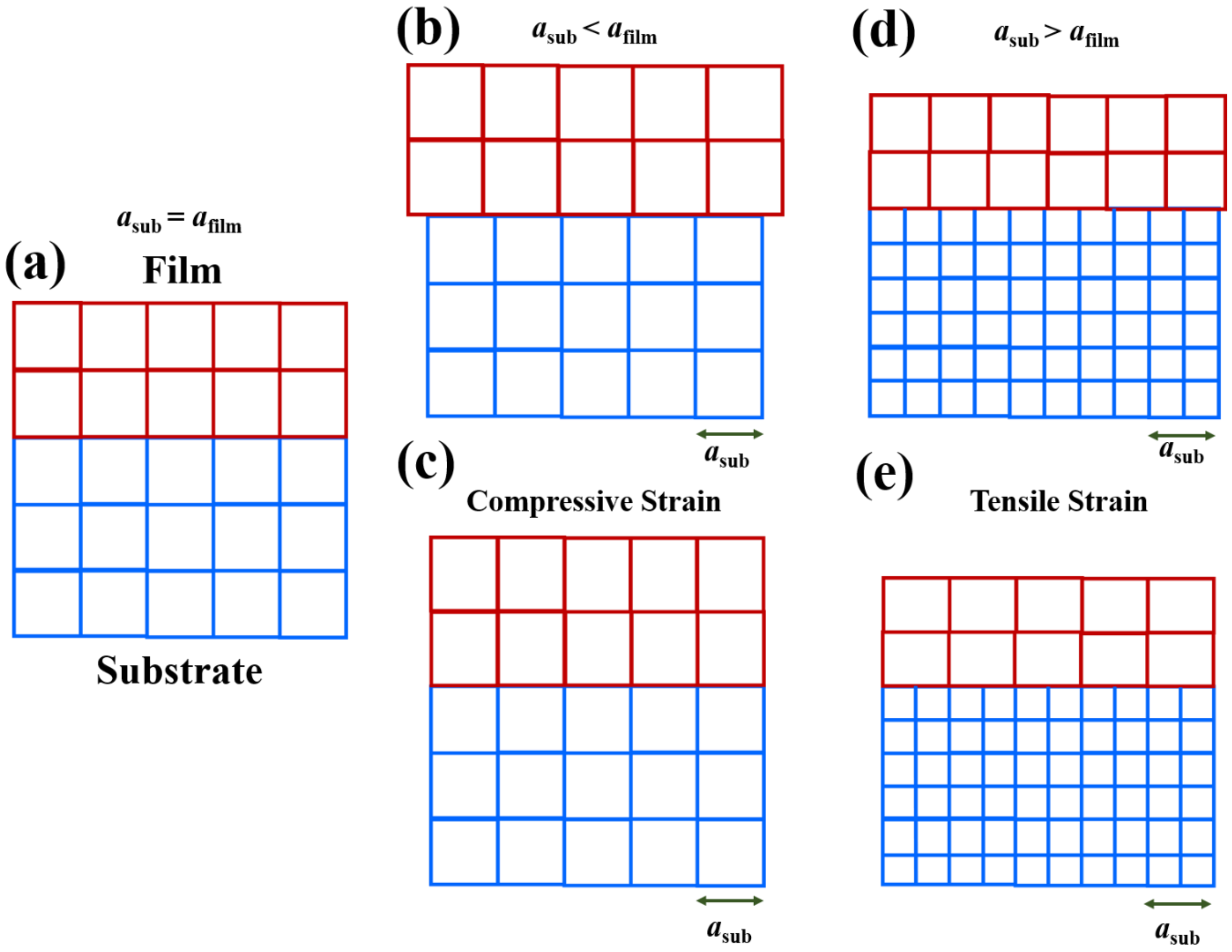

**Fig.1**

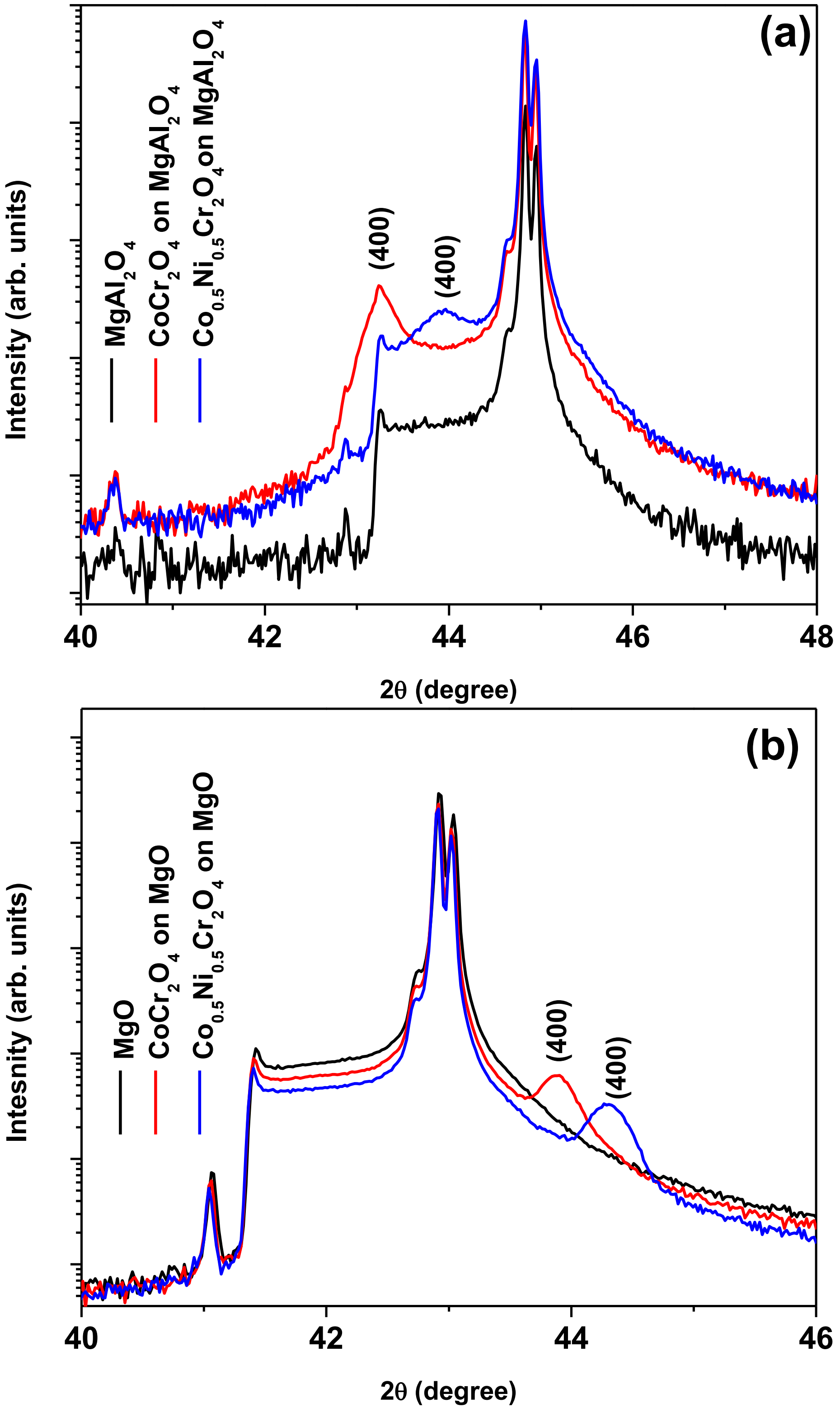

**Fig.2**

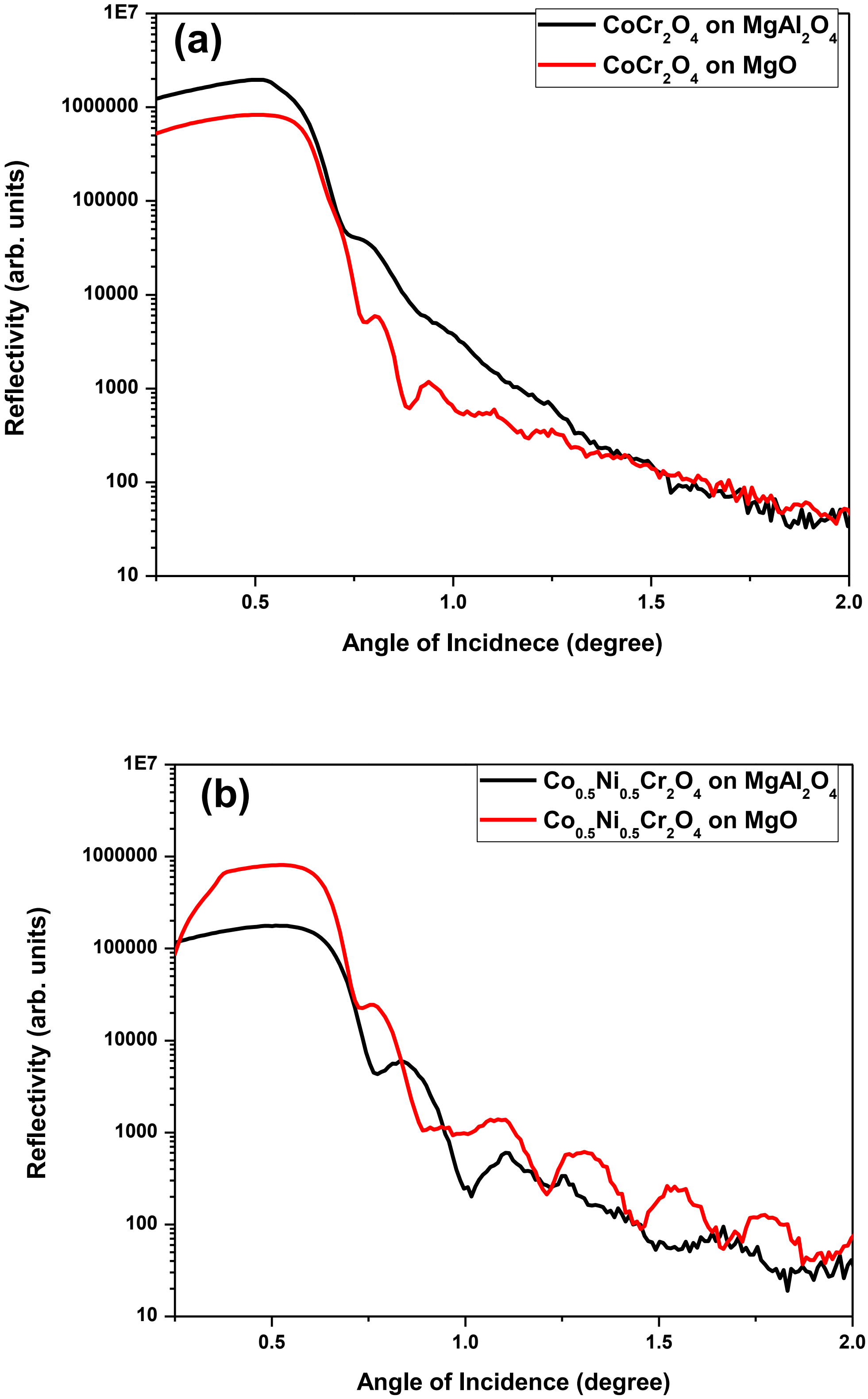

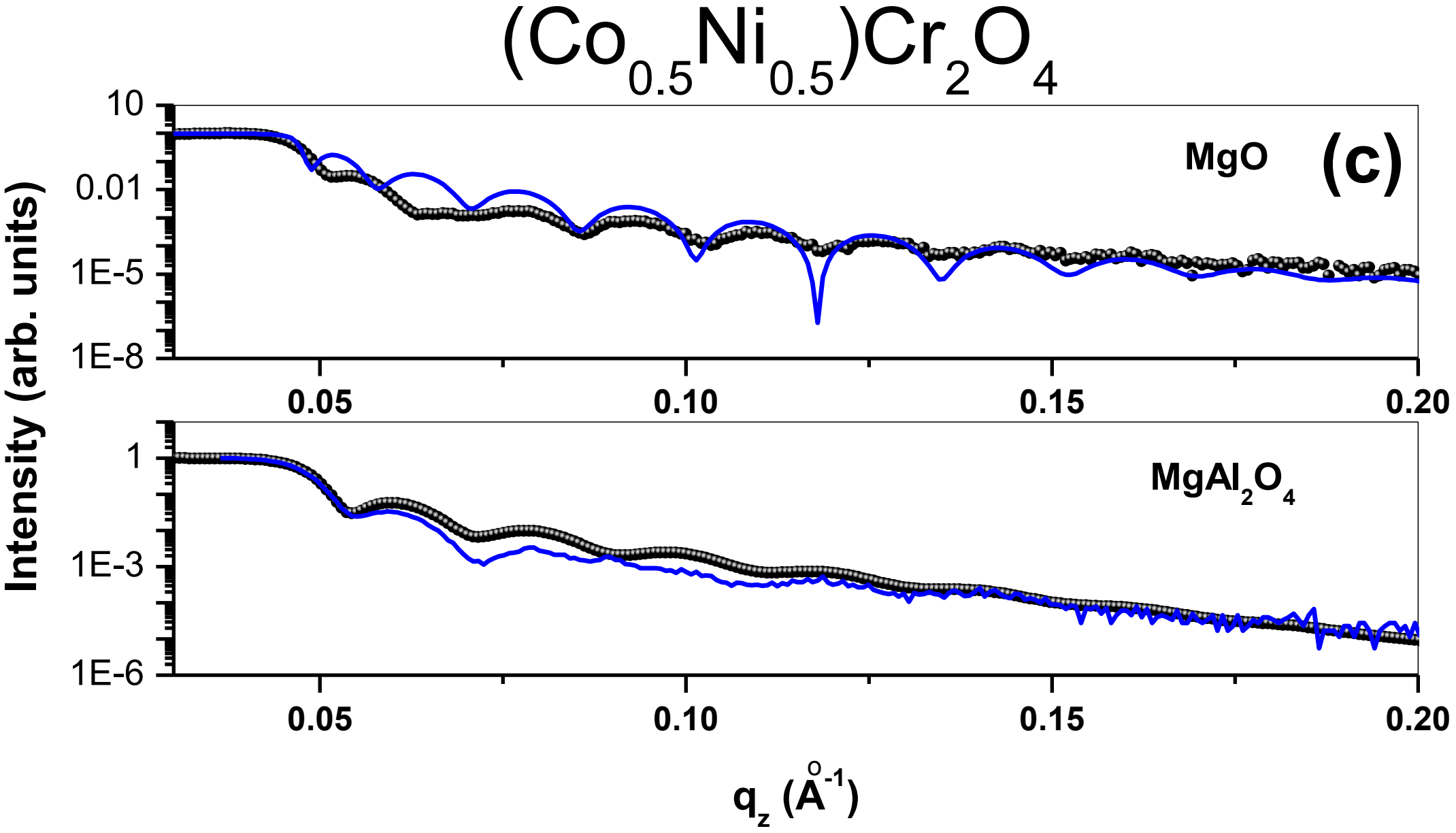

**Fig. 3.**

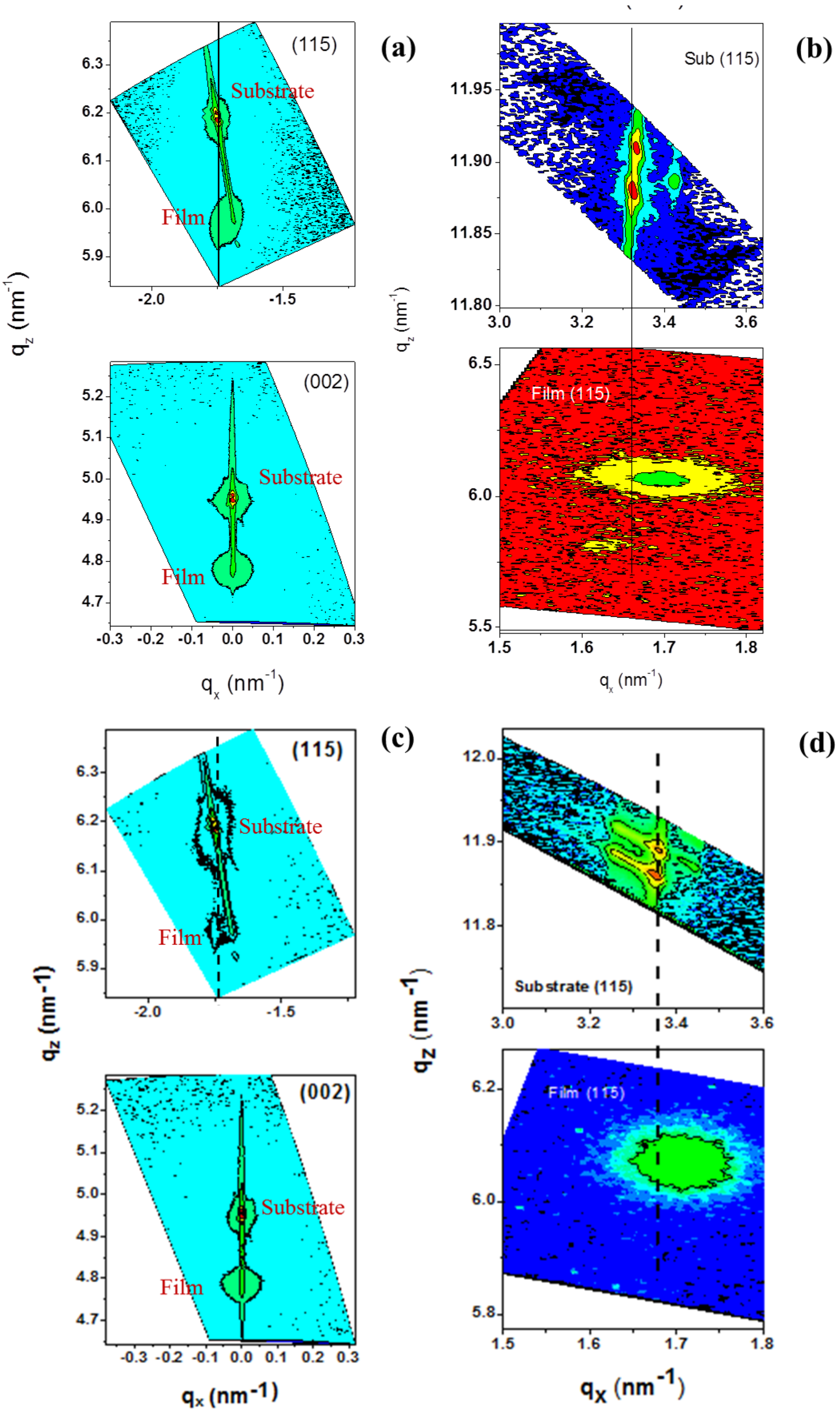

**Fig 4.**

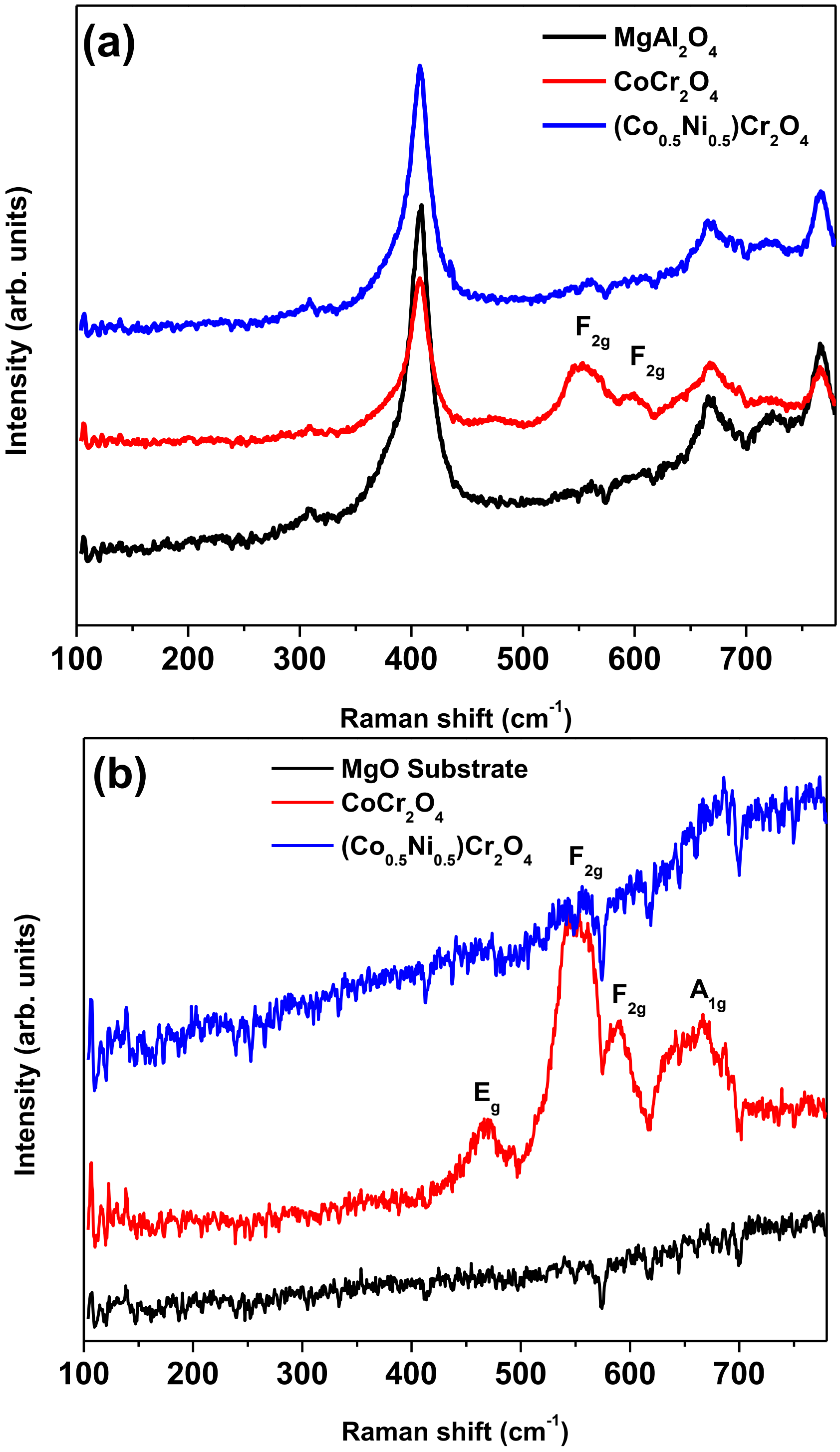

**Fig. 5**

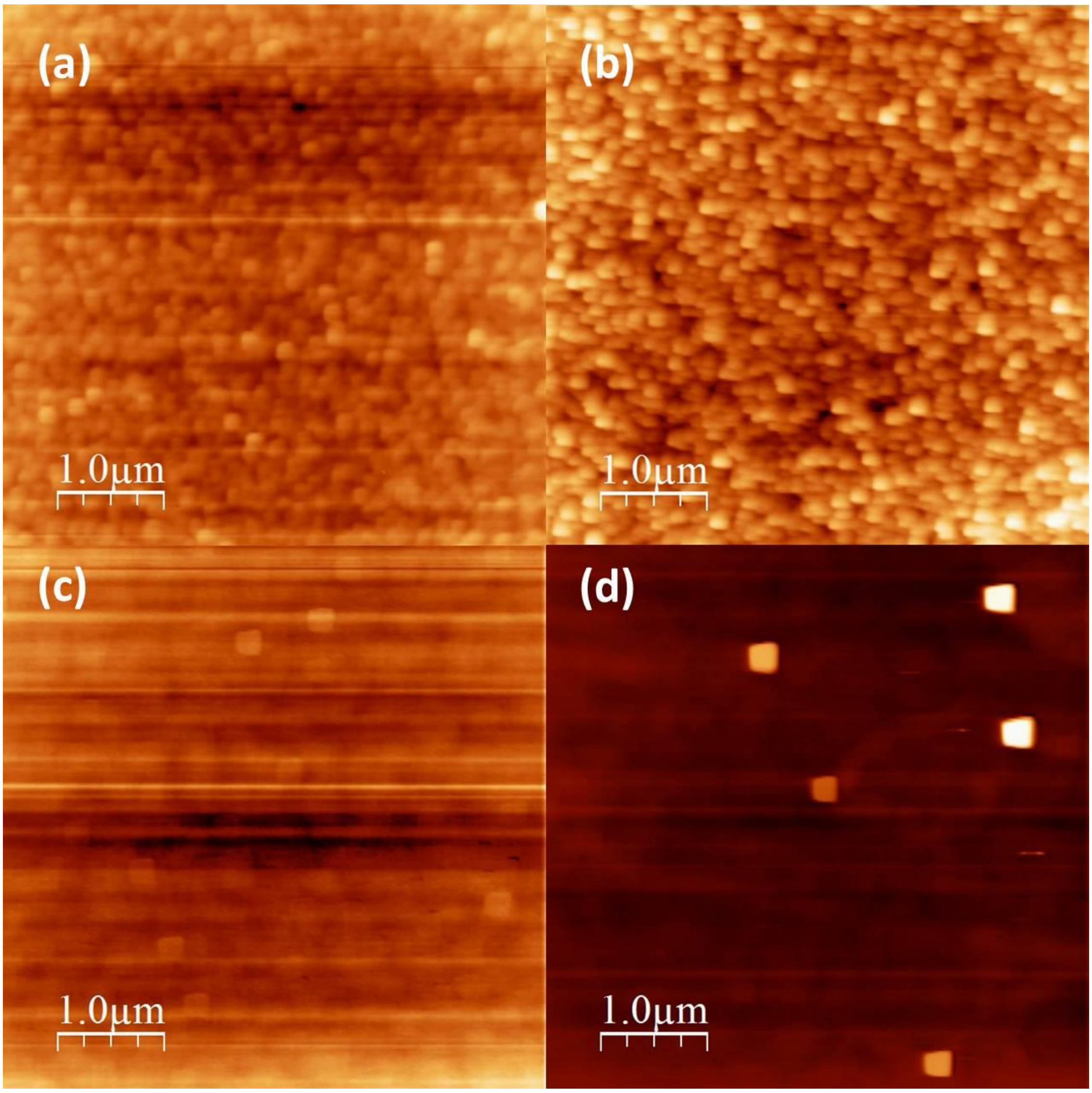

**Fig. 6**

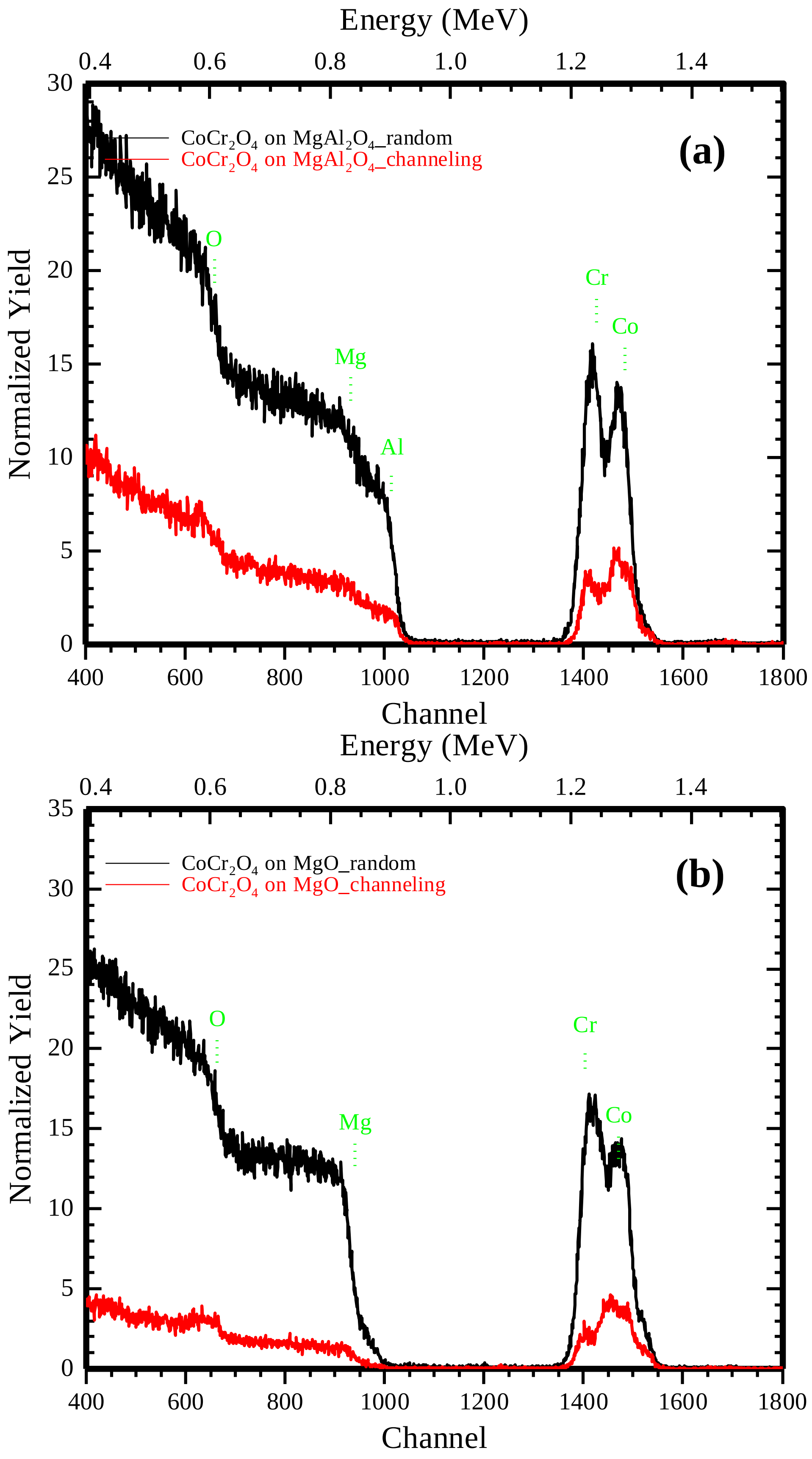

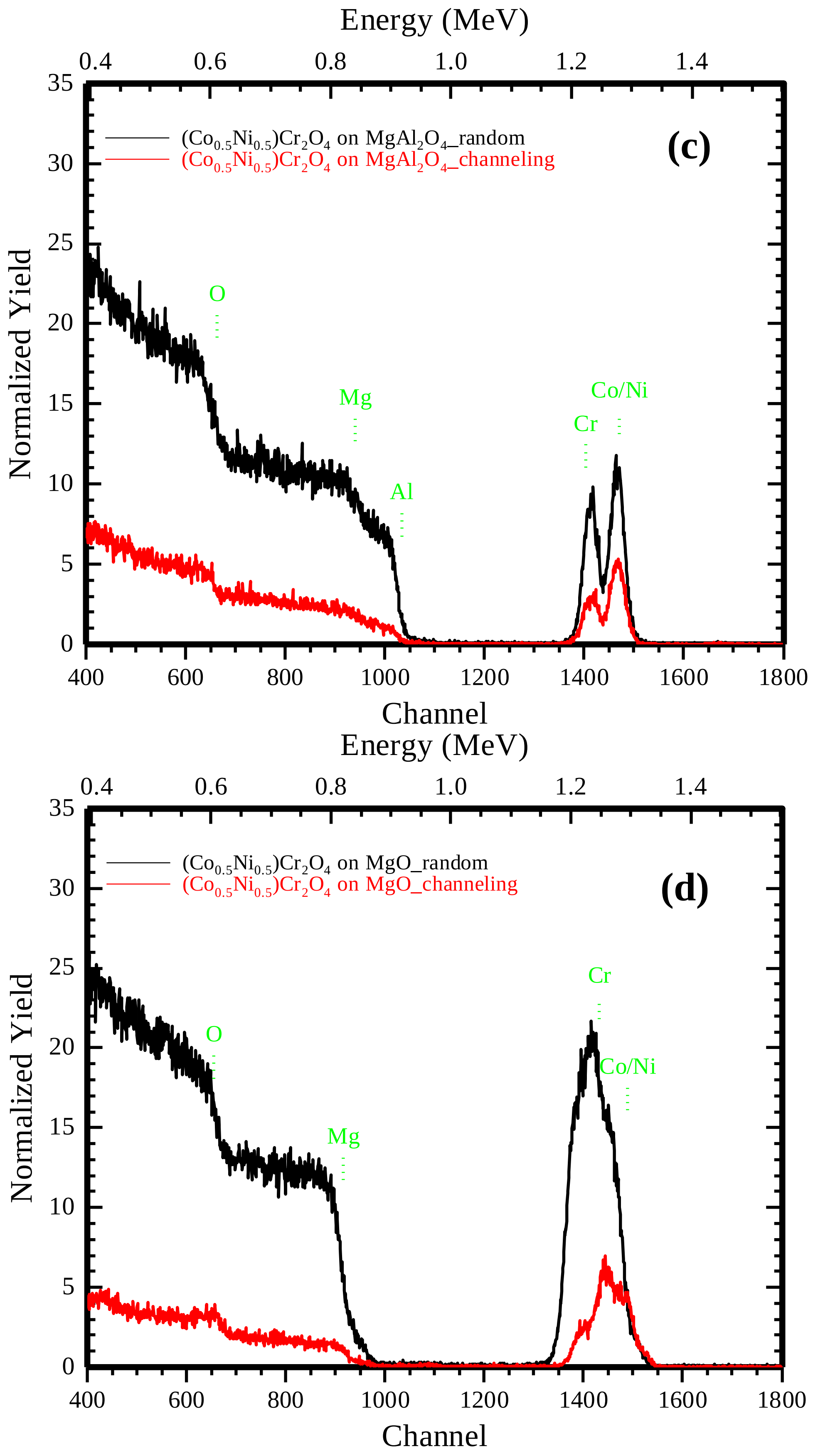

**Fig. 7**

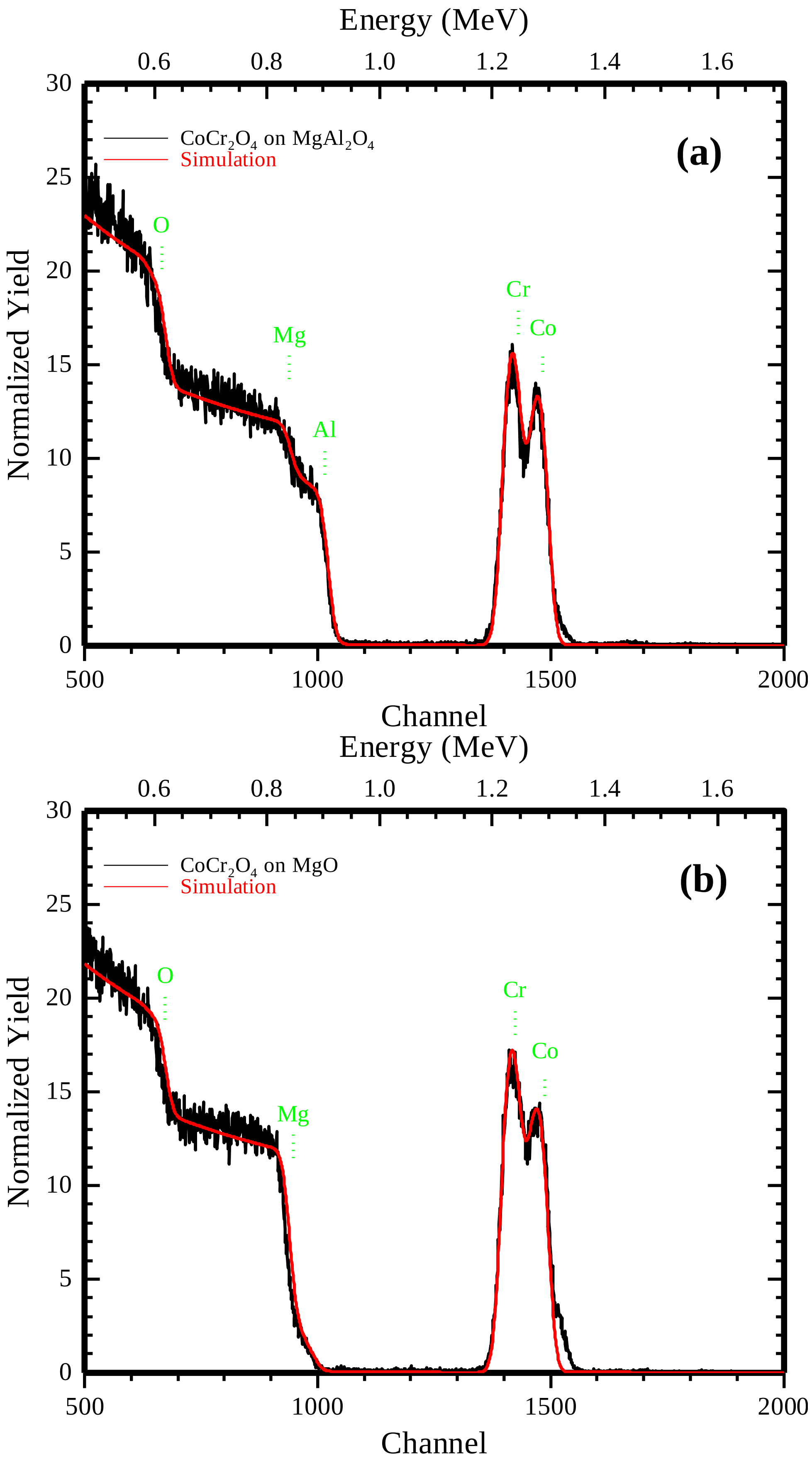

**Fig. 8**

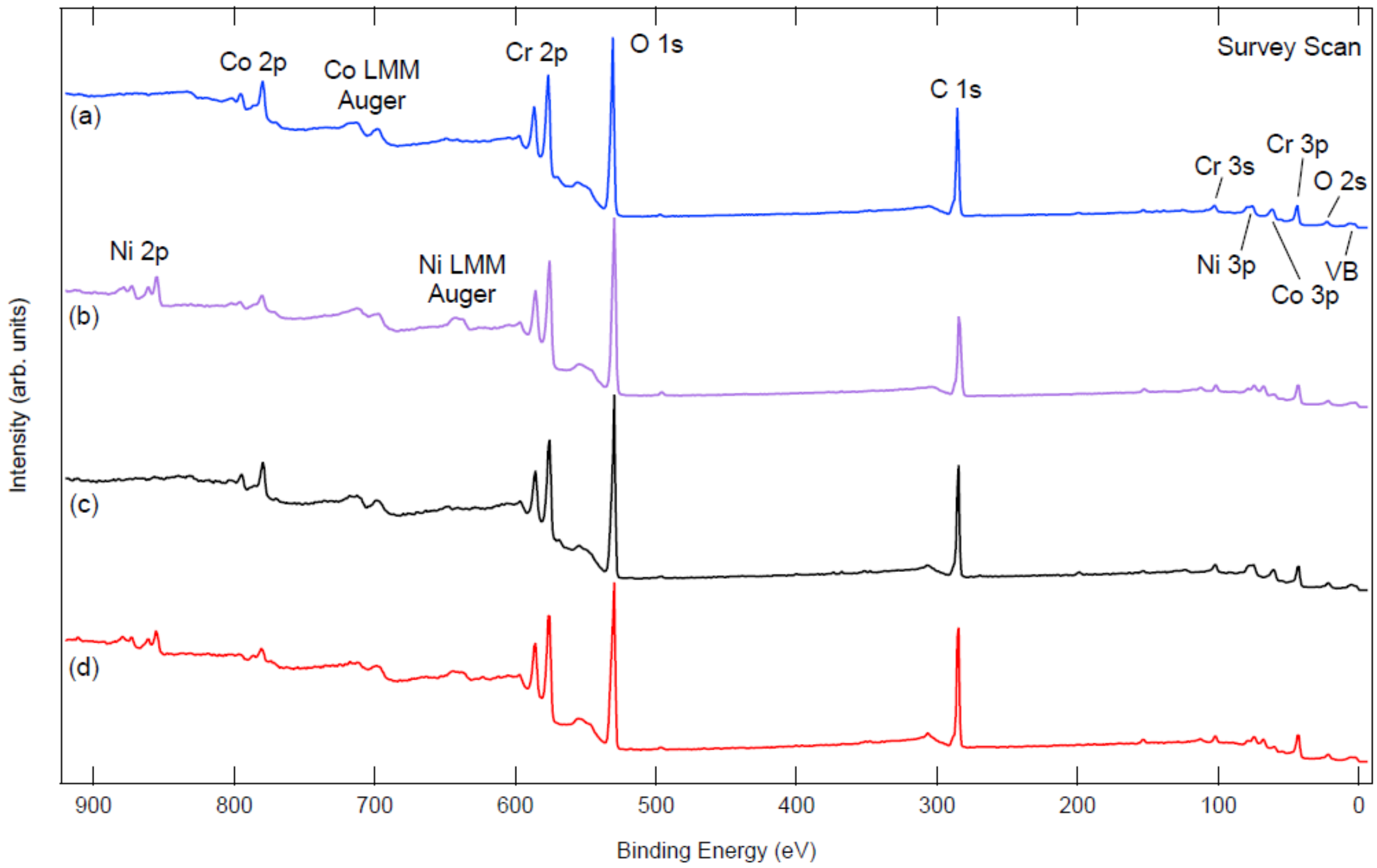

**Fig. 9**

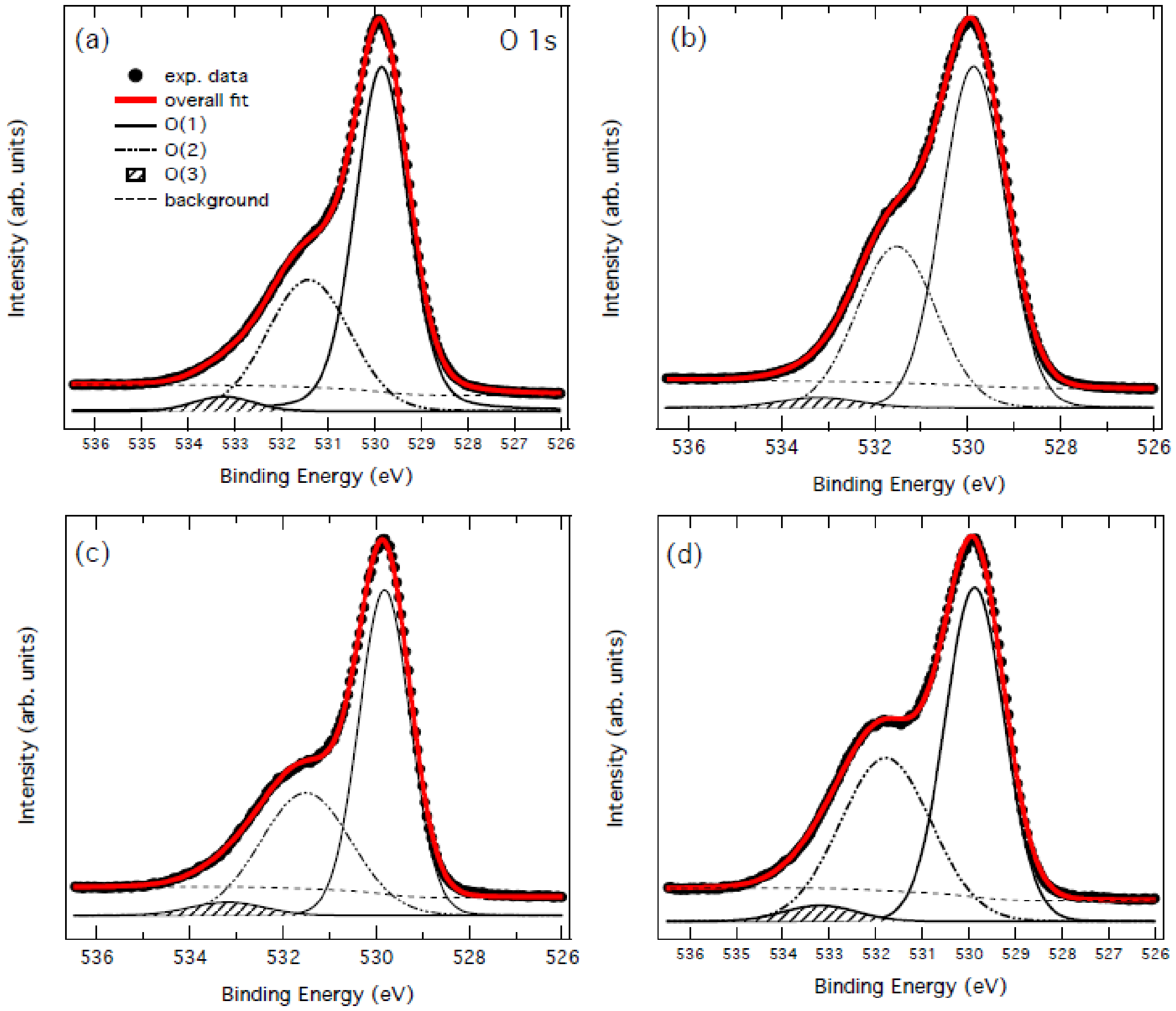

**Fig. 10**

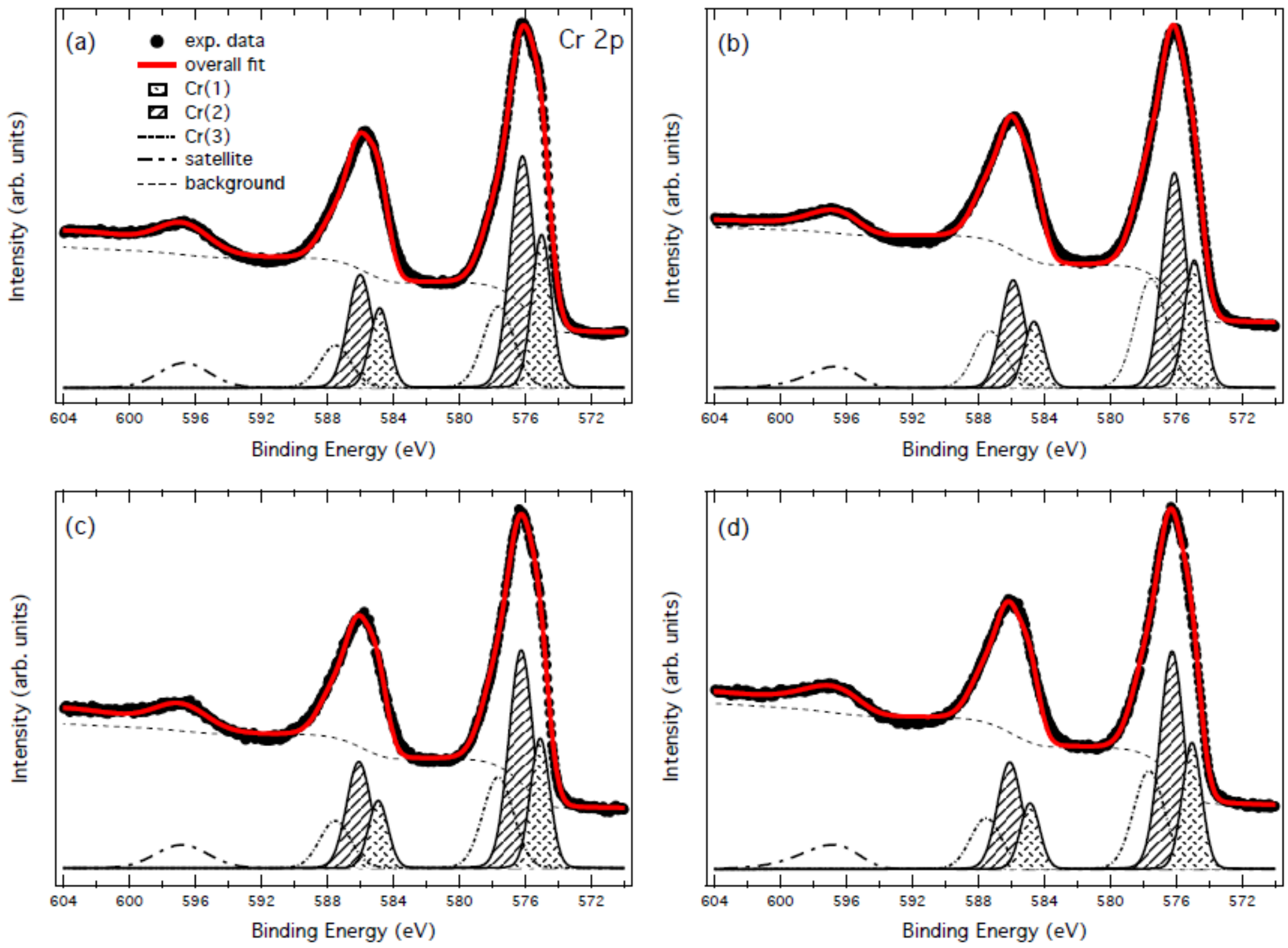

**Fig. 11**

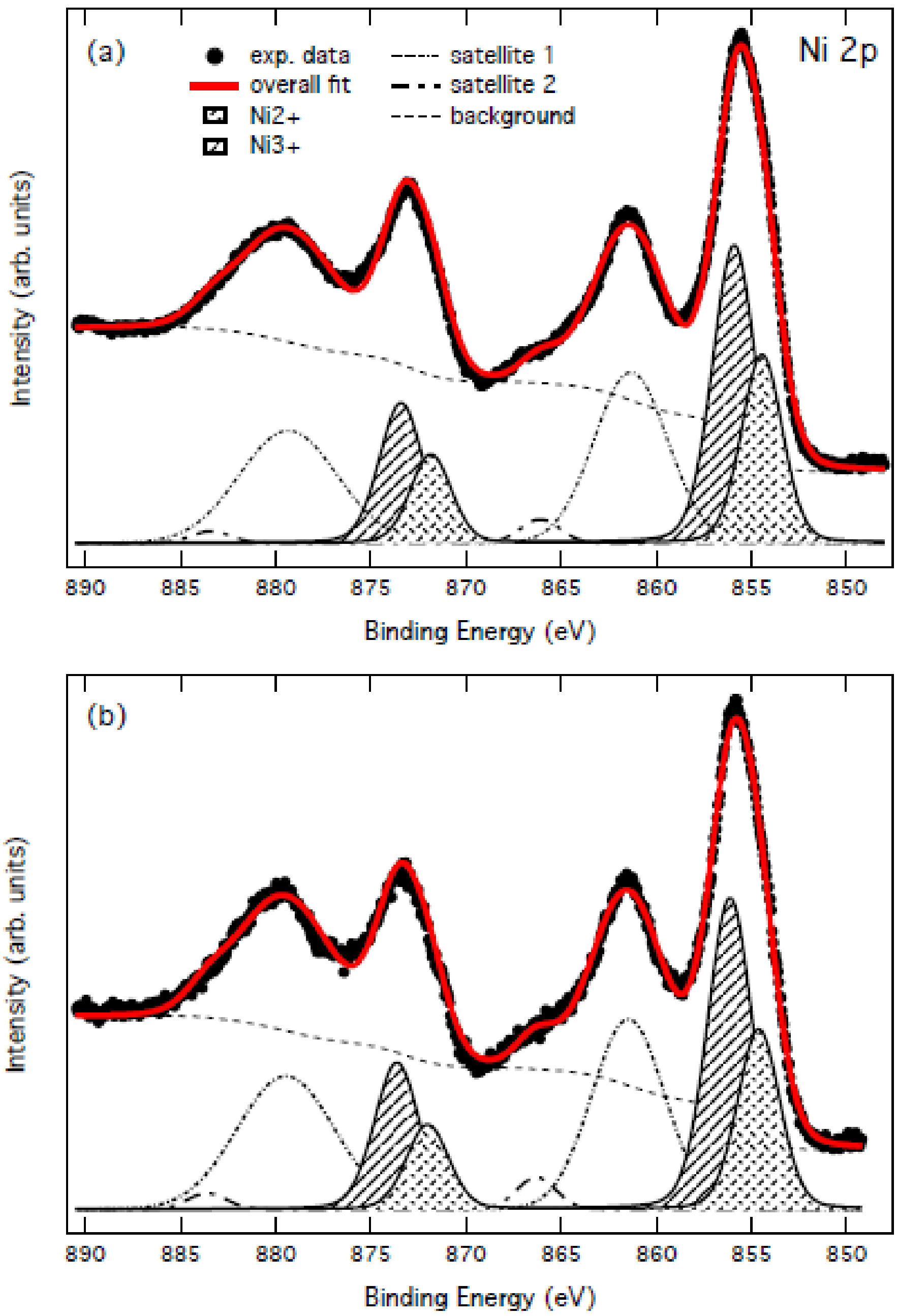

**Fig. 12**

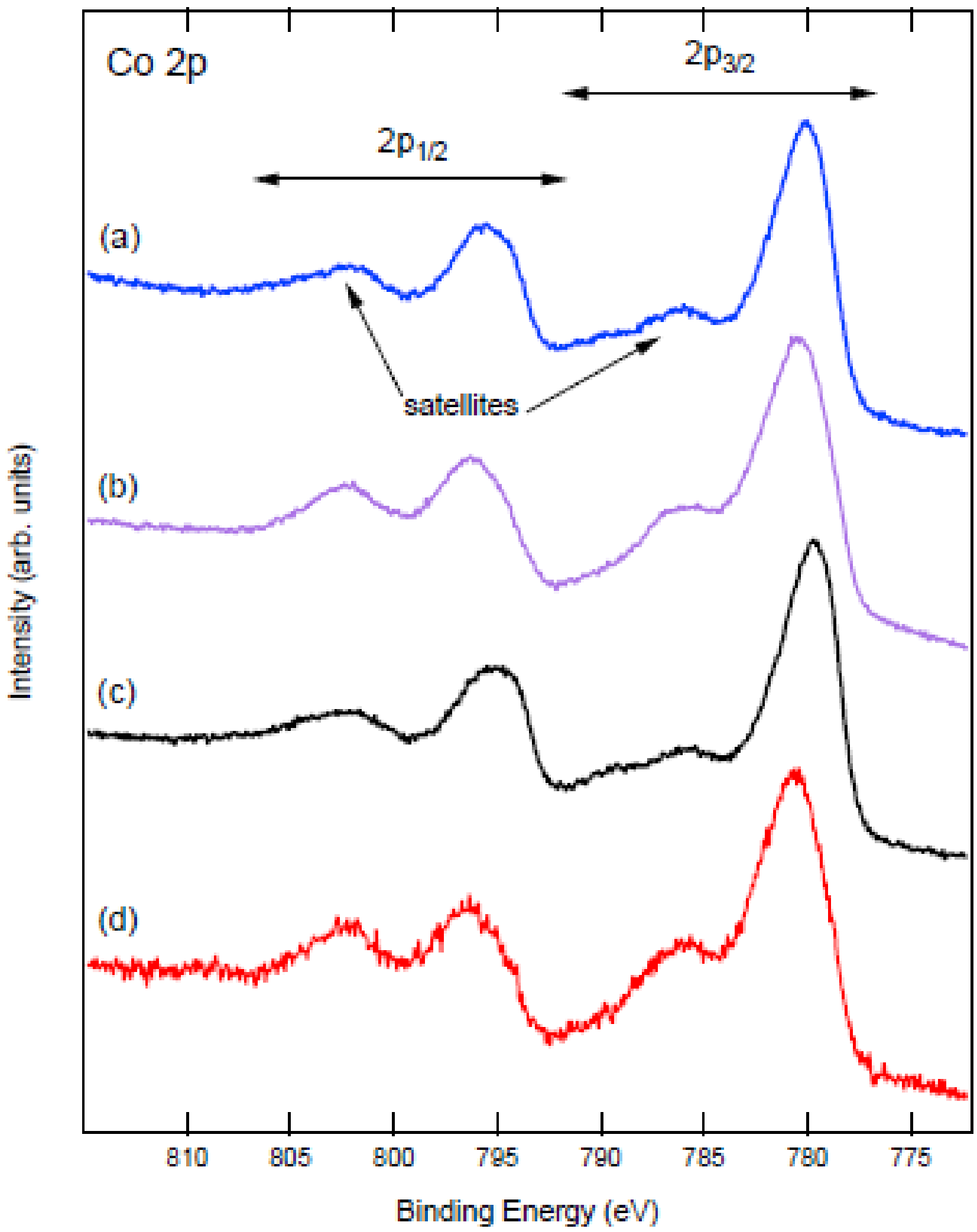

**Fig. 13**

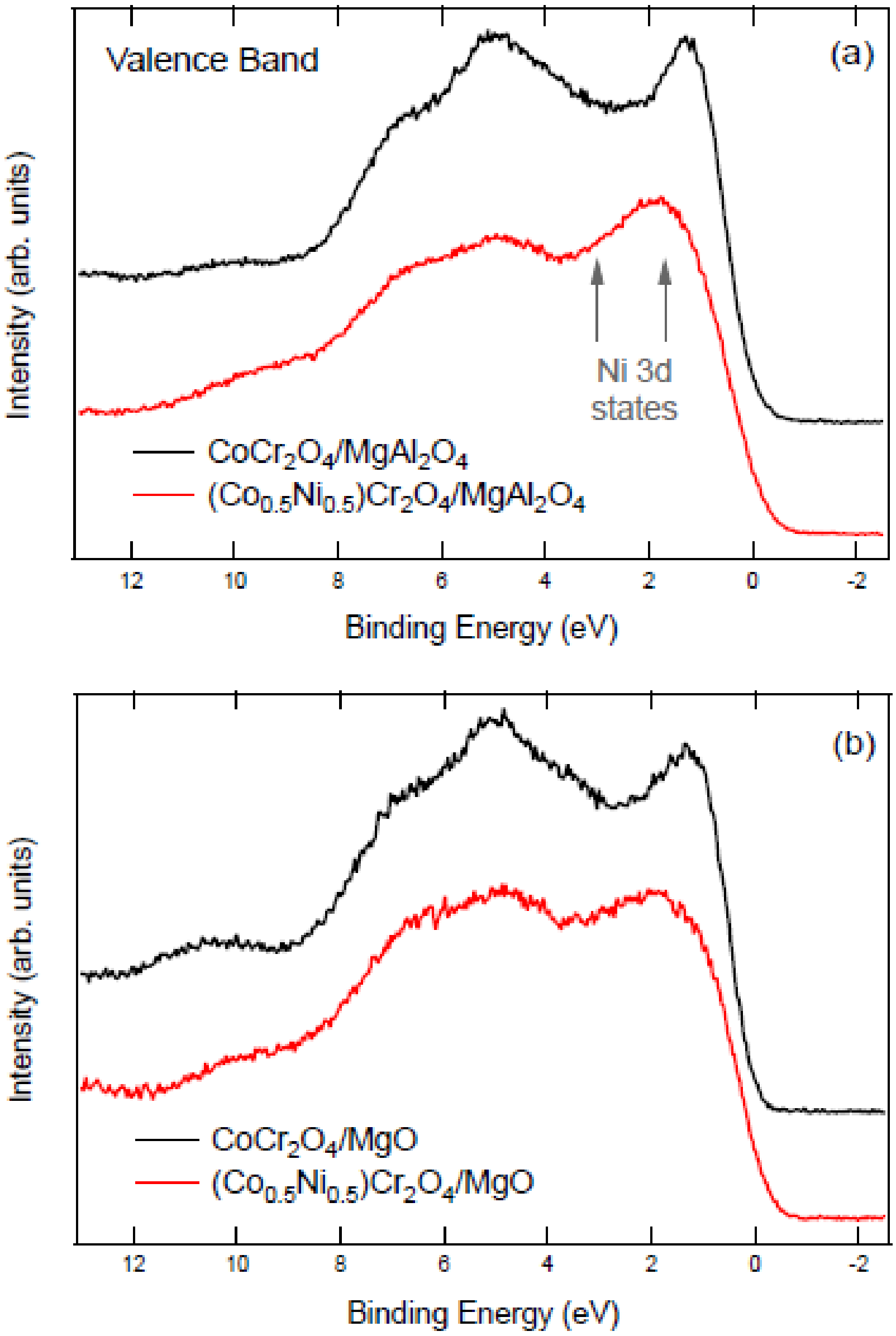

**Fig. 14**

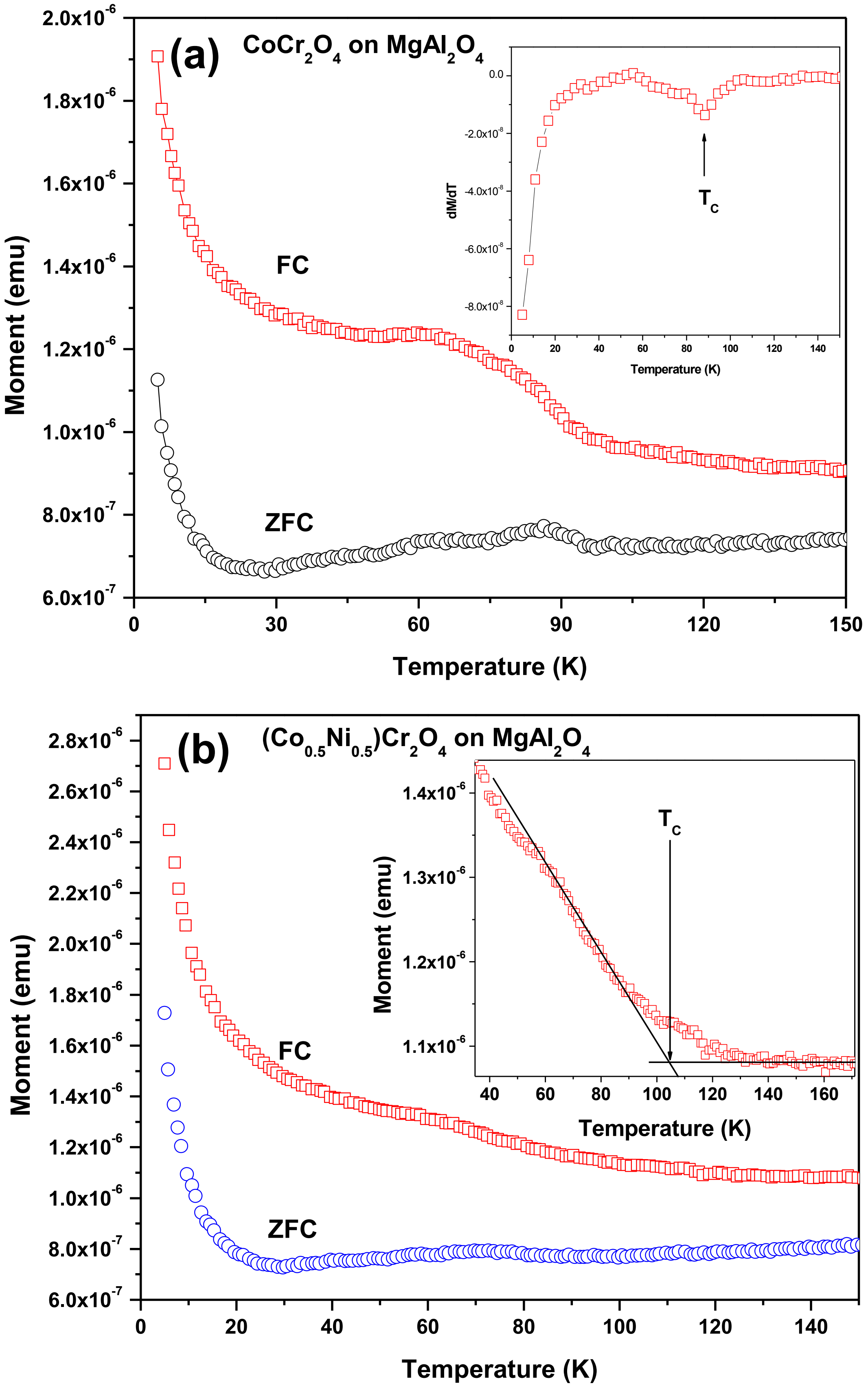

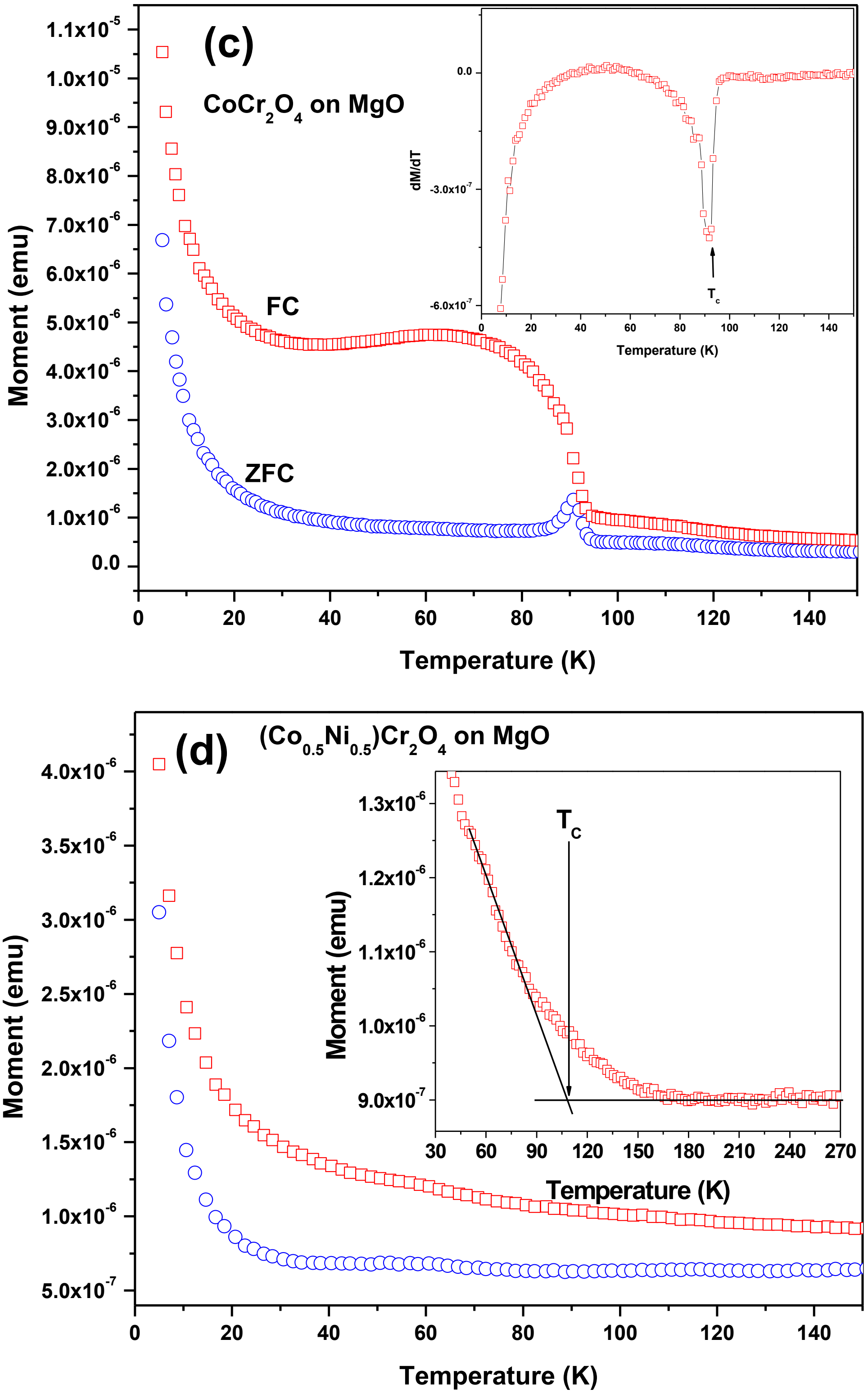

**Fig. 15**

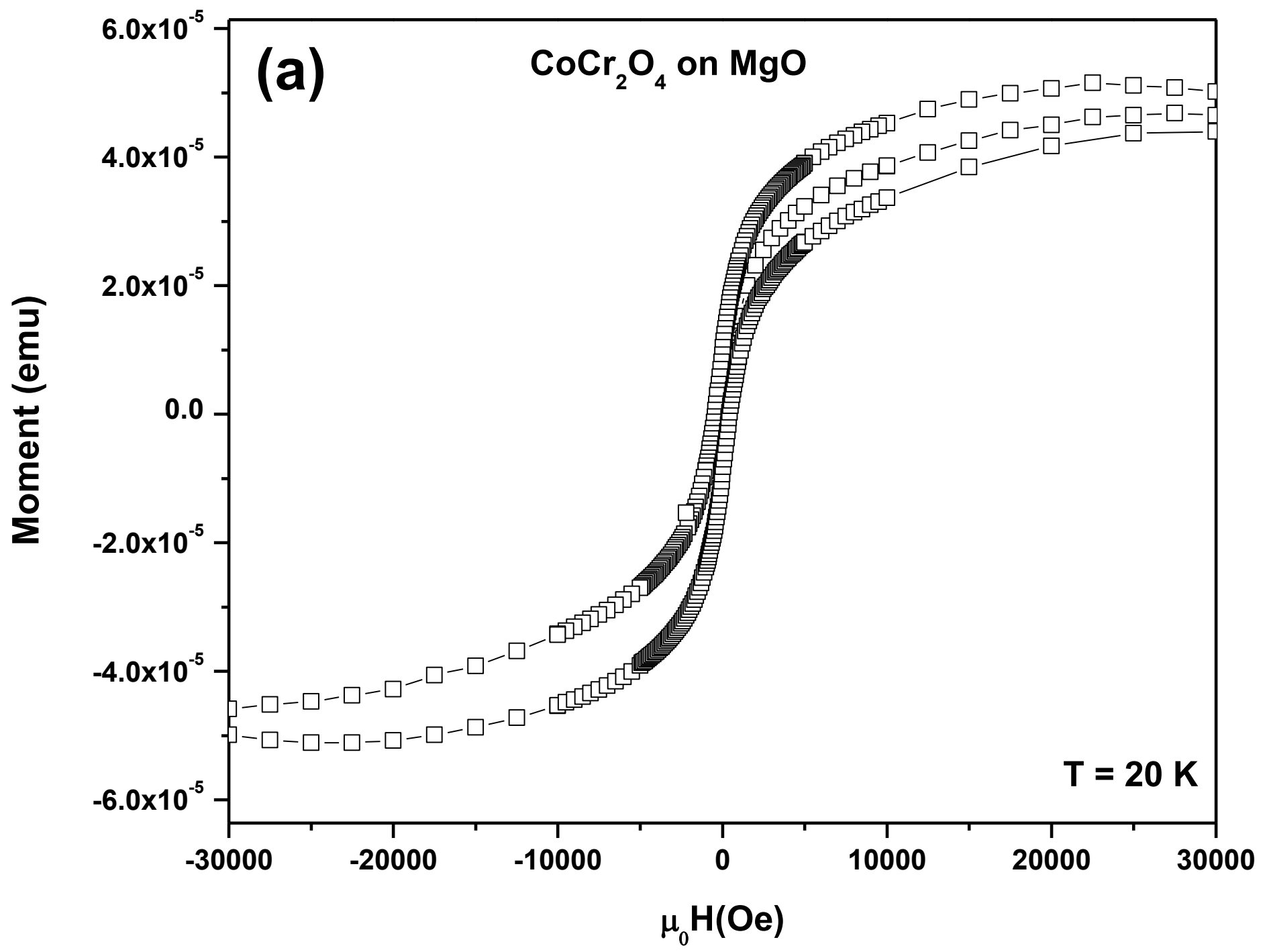

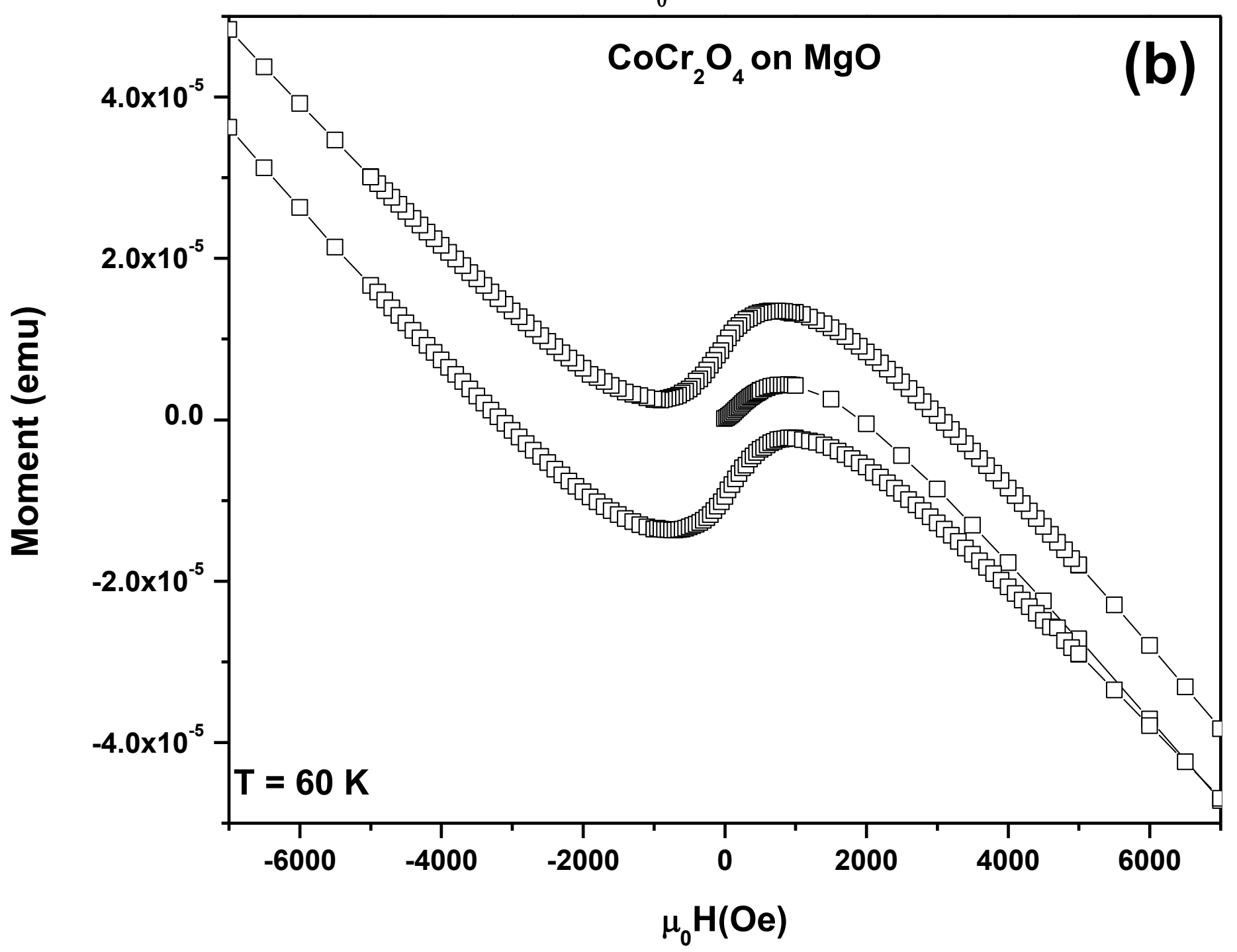

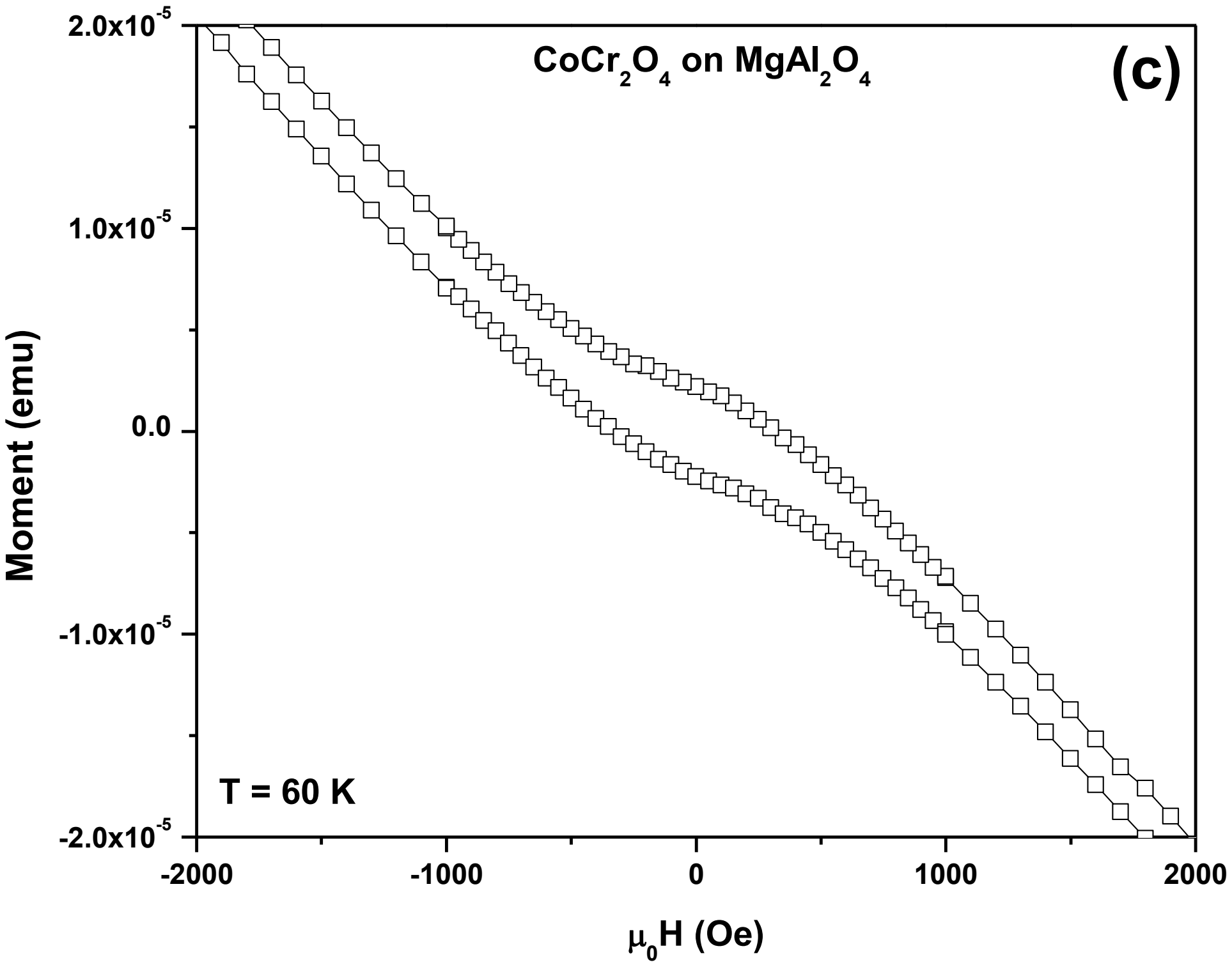

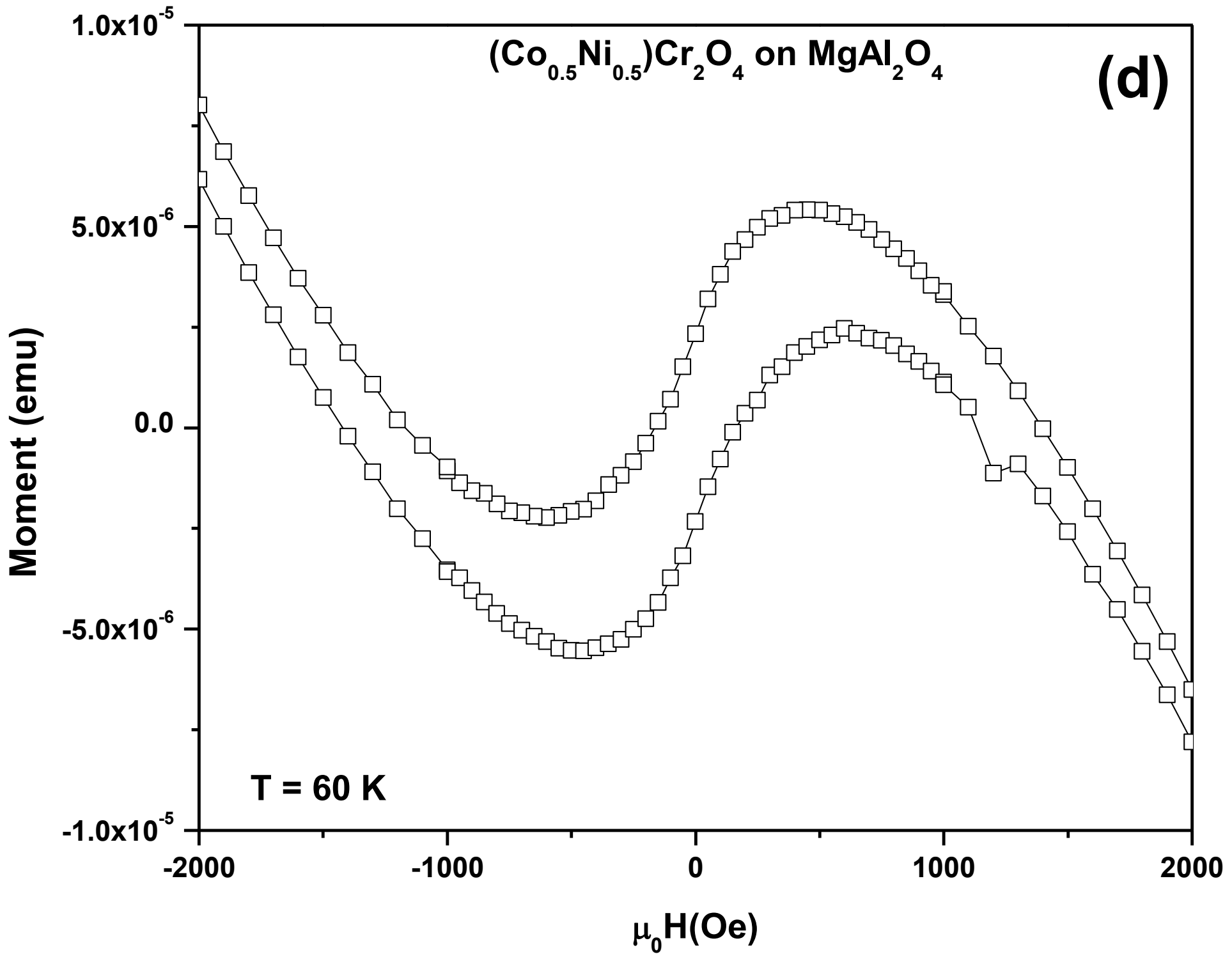

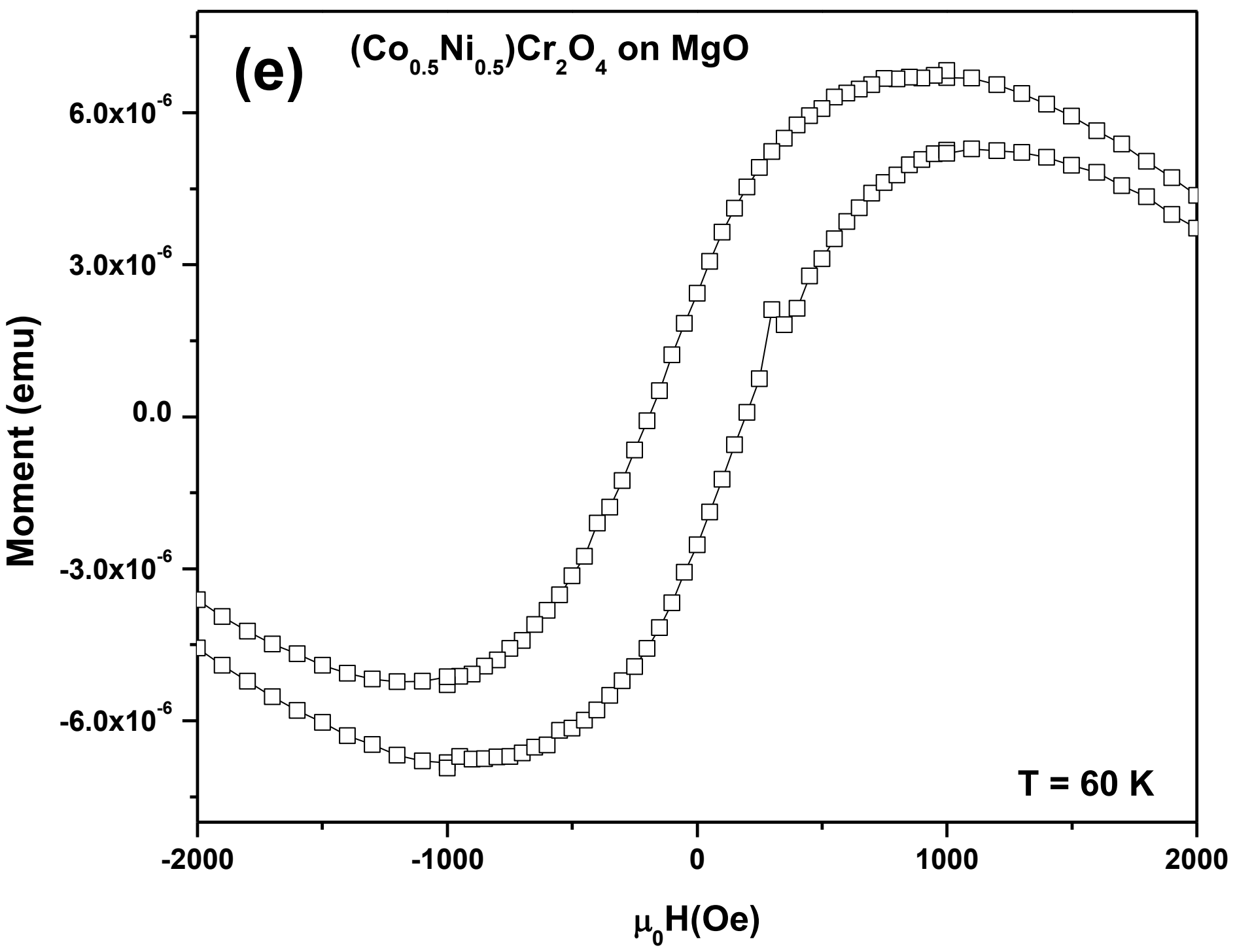

Fig. 16